%% file: ms.tex
\documentclass[iop,twocolappendix]{emulateapj}
\usepackage{amsmath}

\input{newcom}
\newcommand{\obsA}{0204670101}
\newcommand{\obsB}{0670780101}
\newcommand{\Eedge}{\ensuremath{E_\mathrm{edge}}}
\newcommand{\eCX}{\ensuremath{\varepsilon_\mathrm{CX}}}
\newcommand{\lCX}{\ensuremath{l_\mathrm{CX}}}
\newcommand{\ncl}{\ensuremath{n_\mathrm{cl}}}
\newcommand{\nh}{\ensuremath{n_\mathrm{h}}}
\newcommand{\EK}{\ensuremath{E_\mathrm{K}}}
\newcommand{\Lmag}{\ensuremath{L_\mathrm{mag}}}
\newcommand{\tmag}{\ensuremath{t_\mathrm{mag}}}
\newcommand{\csix}{\ensuremath{\mathrm{C}^{+6}}}
\newcommand{\cfive}{\ensuremath{\mathrm{C}^{+5}}}
\newcommand{\nseven}{\ensuremath{\mathrm{N}^{+7}}}
\newcommand{\nsix}{\ensuremath{\mathrm{N}^{+6}}}
\newcommand{\nfive}{\ensuremath{\mathrm{N}^{+5}}}
\newcommand{\oeight}{\ensuremath{\mathrm{O}^{+8}}}
\newcommand{\oseven}{\ensuremath{\mathrm{O}^{+7}}}
\newcommand{\osix}{\ensuremath{\mathrm{O}^{+6}}}

\shorttitle{ORIGIN OF X-RAY EMISSION FROM MS30.7}
\shortauthors{HENLEY ET AL.}

\begin{document}

\title{The Origin of the X-ray Emission from the High-velocity Cloud MS30.7$-$81.4$-$118}
\author{David B. Henley\altaffilmark{1}, Robin L. Shelton\altaffilmark{1}, and Kyujin Kwak\altaffilmark{2}}
\affil{$^1$Department of Physics and Astronomy, University of Georgia, Athens, GA 30602;\\
  dbh@physast.uga.edu, rls@physast.uga.edu \\
  $^2$School of Natural Science, Ulsan National Institute of Science and Technology (UNIST), \\
  50 UNIST-gil, Ulju-gun, Ulsan, Republic of Korea, 689-798; \\
  kkwak@unist.ac.kr}

\begin{abstract}
A soft X-ray enhancement has recently been reported toward the high-velocity cloud
MS30.7$-$81.4$-$118 (MS30.7), a constituent of the Magellanic Stream. In order to investigate the origin
of this enhancement, we have analyzed two overlapping \xmm\ observations of this cloud.
We find that the X-ray enhancement is $\sim$6\arcmin\ or $\sim$100~\pc\ across, and is concentrated
to the north and west of the densest part of the cloud.
We modeled the X-ray enhancement with a variety of spectral models. A single-temperature equilibrium
plasma model yields a temperature of $(3.69^{+0.47}_{-0.44}) \times 10^6~\K$ and a
0.4--2.0~\kev\ luminosity of $7.9 \times 10^{33}~\ergps$. However, this model underpredicts the
on-enhancement emission around 1~\kev, which may indicate the additional presence of hotter plasma
($T \ga 10^7~\K$), or that recombination emission is important.
We examined several different physical models for the origin of the X-ray enhancement. We find that
turbulent mixing of cold cloud material with hot ambient material, compression or shock heating of a
hot ambient medium, and charge exchange reactions between cloud atoms and ions in a hot ambient
medium all lead to emission that is too faint. In addition, shock heating in a cool or warm medium
leads to emission that is too soft (for reasonable cloud speeds). We find that magnetic reconnection
could plausibly power the observed X-ray emission, but resistive magnetohydrodynamical simulations
are needed to test this hypothesis.
If magnetic reconnection is responsible for the X-ray enhancement, the observed spectral properties
could potentially constrain the magnetic field in the vicinity of the Magellanic Stream.
\end{abstract}


\section{INTRODUCTION}
\label{sec:Introduction}

\setcounter{footnote}{2}

High-velocity clouds (HVCs) are clouds in the Galactic halo with high line-of-sight velocities
($\ga$90~\kmps) relative to the Local Standard of Rest \citep{wakker97}. These clouds may be
condensations from a Galactic fountain falling back toward the disk, gas stripped off satellite
galaxies, or extragalactic gas left over from the formation of the Local Group galaxies (see
\citealt{bregman04} for a review). Note that not all HVCs need have the same origin, and more than
one of the aforementioned processes may have been involved in the creation of the Galaxy's
population of HVCs. HVCs were originally discovered from observations of 21-cm \HI\ emission
\citep{muller63}, and a recent survey of moderate and high Galactic latitudes found high-velocity
\HI\ on 37\%\ of sight lines \citep{lockman02}. However, high-velocity material is also observed via
other lines, including those of high ions such as \CIV, \NV, and
\OVI\ \citep{sembach03,fox04,fox05,fox06,collins07}. The high-velocity material bearing these high
ions ($T \sim (\mbox{1--3}) \times 10^5~\K$) is likely produced from the interactions of HVCs with a
hot ($T \ga 10^6~\K$) ambient medium \citep{sembach03,fox04,collins07,kwak11}.

The interactions of HVCs with their surroundings may also produce soft X-ray emission. Several
authors have reported excess soft X-ray emission (above the level of the diffuse X-ray background)
associated with some HVCs. The early reports were based on Wisconsin, \textit{SAS-3}, and
\textit{HEAO-1} survey data of the Complex~C region \citep{hirth85}, a pointed \rosat\ observation
of HVC~90.5$+$42.5$-$130 \citep{kerp94}, and \rosat\ All-Sky Survey data of several different HVC
complexes \citep{herbstmeier95,kerp96,kerp99}. For the HVC complexes, their reported X-ray excesses
are $\sim$1\degr--10\degr\ in size \citep{herbstmeier95,kerp96,kerp99}. However, the conclusion that
these X-ray excesses are physically associated with the HVCs has been disputed, on the grounds that
these studies did not adequately take into account the possibility of small-scale variations in the
brightnesses of the Galactic halo and/or the Local Bubble \citep{wakker97,wakker99a}. Better data at
lower energies (where the absorption is higher) are needed to constrain the location of these X-ray
excesses relative to the Galaxy's \HI\ \citep{wakker99a}.

More recently, higher spatial resolution observations with \xmm\ and \chandra\ show evidence for
excess X-ray emission (on the scale of a few arcminutes) associated with other, more compact HVCs
\citep{bregman09a}.\footnote{Note that the \xmm\ and \chandra\ data are also of higher spectral
  resolution than the earlier data. However, \citet{bregman09a} did not report on the spectra of the
  X-ray excesses that they observed.} The small spatial scales compared with the
above-mentioned studies increase the confidence that the associations between these X-ray excesses
and the corresponding HVCs are real. The best evidence comes from an \xmm\ observation of the HVC
MS30.7$-$81.4$-$118 (hereafter MS30.7), which is a constituent of the Magellanic Stream
\citep{mathewson74,putman03}. \citet{bregman09a} were looking for shadowing (i.e., partial blocking)
of the X-ray emission expected from the warm-hot intergalactic medium (WHIM; \citealt{cen99}), and
thus hoped to see a reduction in the X-ray count rate toward the densest part of the cloud. Instead,
they found that 0.4--1.0~\kev\ count rate measured with the pn detector toward the densest part of
the cloud was $0.64 \pm 0.10~\counts\ \pks\ \parcminsq$ higher than the off-cloud background rate
(2.54 versus 1.90~counts \pks\ \parcminsq). \citet{bregman09a} also found on-cloud X-ray excesses
(albeit less significant) in \chandra\ observations of MS30.7 and of another HVC,
CHVC~125$+$41$-$207. They attributed the greater significance of the enhancement in the
\xmm\ observation of MS30.7 to \xmm's larger field of view, which allowed for greater contrast
between the on- and off-cloud regions.

If the observed X-ray excesses are physically associated with HVCs, then they undoubtedly arise from
the interaction of the clouds with their environment. However, the detailed mechanism is
uncertain. Various mechanisms have been proposed: shock heating
\citep{hirth85,bregman09a,shelton12}, compression of an already hot ambient medium
\citep{herbstmeier95}, magnetic reconnection \citep{kerp94,kerp96,zimmer97}, charge exchange
reactions between neutral atoms in the cloud and ions in a hot ambient medium \citep{lallement04b},
and turbulent mixing of the cold cloud material with a hot ambient medium
\citep{shelton12}. Regardless of the specific mechanism, the resulting X-ray spectrum and brightness
will depend, at least in part, on the ambient conditions in the vicinity of the HVC. As a result,
understanding the mechanism behind the X-ray emission from HVCs could potentially provide
constraints on quantities such as the density, the pressure, or the magnetic field in the Galactic
halo. Note that, at the distance of the Magellanic Stream ($\sim$60~\kpc), few such constraints
exist.

In this paper, we analyze two \xmm\ observations of MS30.7, with the goal of understanding the
mechanism responsible for the X-ray enhancement reported by \citet{bregman09a}. The first
observation is that analyzed by \citeauthor{bregman09a}, while the second is to the east of the
first (Section~\ref{sec:Observations}).  We use these observations to create an X-ray image of
MS30.7 and its surroundings, so we can more clearly see the extent and morphology of the X-ray
enhancement (Section~\ref{subsec:ImageCreation}). We extract
(Section~\ref{subsec:SpectralExtraction}) and analyze (Section~\ref{sec:Analysis}) the spectrum of
the X-ray enhancement, using a variety of spectral models. We then use these results to test
different physical models for the origin of the X-ray emission (Section~\ref{sec:Models}).  In
particular, we examine turbulent mixing with or compression of a hot ambient medium
(Section~\ref{subsec:TurbulentMixing}), shock heating (Section~\ref{subsec:ShockHeating}), charge
exchange (Section~\ref{subsec:ChargeExchange}), and magnetic reconnection
(Section~\ref{subsec:MagneticReconnection}). We discuss our results in Section~\ref{sec:Discussion},
and finish with our summary and conclusions in Section~\ref{sec:Summary}.

\section{OBSERVATIONS AND DATA REDUCTION}
\label{sec:Observations}

The details of the two \xmm\ observations of MS30.7 are shown in Table~\ref{tab:Obs}.  The first
observation (observation ID \obsA) was previously analyzed by \citet{bregman09a}. The second
observation (observation ID \obsB), carried out 7.5~years later, has its pointing direction
$\approx$14\arcmin\ to the east of that of the first observation. Since the radius of the
\xmm\ field of view is $\approx$14\arcmin, the fields overlap.

\begin{deluxetable*}{cccccccc}
\tablecaption{MS30.7 Observation Details\label{tab:Obs}}
\tablehead{
\colhead{Obs.\ ID} & \colhead{Start date} & \colhead{R.A.\ (J2000)}  & \colhead{Dec.\ (J2000)} & \colhead{$t_\mathrm{exp}$ (ks)} & \multicolumn{3}{c}{$t_\mathrm{clean}$ (ks)} \\
\cline{6-8}
                   &                      &                          &                         &                               & MOS1          & MOS2          & pn \\
\colhead{(1)}      & \colhead{(2)}        & \colhead{(3)}            & \colhead{(4)}           & \colhead{(5)}                 & \colhead{(6)} & \colhead{(7)} & \colhead{(8)}
}
\startdata
\obsA              & 2004 Jan 03          & 00 12 56.3               & $-27$ 12 06.7           & 51.9                          &  44.1         & 43.7          & 35.6 \\
\obsB              & 2011 Jul 01          & 00 14 08.7               & $-27$ 11 01.2           & 61.5                          & 33.1          & 35.9          & 22.0
\enddata
\tablecomments{Column~(5) contains the observation exposure time, while columns (6)--(8) contain the usable exposure times for each camera after cleaning.}
\end{deluxetable*}

\subsection{Initial Data Reduction}
\label{subsec:InitialDataReduction}

We reduced the data using the \xmm\ Extended Source Analysis
Software\footnote{http://heasarc.gsfc.nasa.gov/docs/xmm/xmmhp\_xmmesas.html} (\esas;
\citealt{snowden12}), as distributed with version 12.0.1 of the \xmm\ Science Analysis
Software\footnote{http://xmm.esac.esa.int/sas/} (SAS). We first used the standard SAS
\texttt{emchain} and \texttt{epchain} scripts to produce calibrated events list for each exposure
(i.e., for each camera from each observation). We then used the \esas\ \texttt{mos-filter} and
\texttt{pn-filter} scripts to excise periods of soft-proton flaring from the data (essentially,
these scripts remove from the data periods of time whose count rates differ by more than $1.5\sigma$
from the typical count rate). The usable time that remained after this cleaning is shown for each
camera in the final three columns of Table~\ref{tab:Obs}.

Note that the soft-proton filtering yields systematically less time for the pn camera than for the
MOS cameras. This is most likely due to the fact that the pn camera is more sensitive than the MOS
cameras. As a result, in the absence of any flaring (i.e., if the count rate variations were due
solely to Poissonian fluctuations), the count rate distribution would be narrower, relative to the
mean count rate, for the pn camera than for the MOS cameras. This means that relatively smaller
excursions from the mean count rate would be flagged as soft proton flares in the pn data than in
the MOS data, resulting in more time being filtered out of the pn data.

We next used the SAS \texttt{edetect\_chain} script to detect sources whose 0.5--2.0~\kev\ flux
exceeded $2 \times 10^{-15}~\flux$. We analyzed each observation individually, using the data from
the MOS1, MOS2, and pn cameras simultaneously. We excluded the detected sources from the data using
circular exclusion regions. For a given source, the source exclusion radius was equal to the
semimajor axis of the ellipse on which the source count rate per pixel is 0.2 times the local
background count rate.  This radius depends on the source brightness relative to the local
background. The source exclusion regions for the two observations were merged before the sources
were excised. Hence, if a source lying in the overlap region was detected in both observations, it
was excised using the larger of the two source exclusion regions resulting from the analysis of the
individual observations. Also, if a source in the overlap region was detected in only one
observation, the region surrounding it was excised from both observations. This approach of merging
the source exclusion regions is conservative in terms of minimizing contamination of the diffuse
emission from point sources.

\subsection{X-ray Image Creation}
\label{subsec:ImageCreation}

We used \esas\ tools to create a mosaicked 0.4--1.2~\kev\ image of MS30.7. The upper limit of this
energy band was chosen to maximize the width of the band while avoiding contamination from the
instrumental Al fluorescence line at 1.49~\kev. We first used the \texttt{mos\_back} and
\texttt{pn\_back} programs to generate images of the 0.4--1.2~\kev\ quiescent particle background
(QPB) for each exposure. These programs use databases of filter-wheel-closed data to construct the
QPB image; these data were scaled to our observations using data from the unexposed corner pixels
that lie outside the field of view \citep{kuntz08a}. We then used the \texttt{merge\_comp\_xmm}
program to combine the images of the 0.4--1.2~\kev\ events from all three cameras and from both
observations (i.e., a total of six images were combined to make the resulting merged
image). Similarly, we used \texttt{merge\_comp\_xmm} to combine the QPB images and the exposure
maps. Finally, we subtracted the combined QPB image from the combined events image, divided this
background-subtracted image by the exposure map, and adaptively smoothed the resulting flat-fielded
image, using the \esas\ \texttt{adapt\_2000} program. This program also filled in the chip gaps and
the holes in the data resulting from the point source removal, using data from neighboring pixels.

\begin{figure*}[t]
  \centering
  \includegraphics[width=0.58\linewidth]{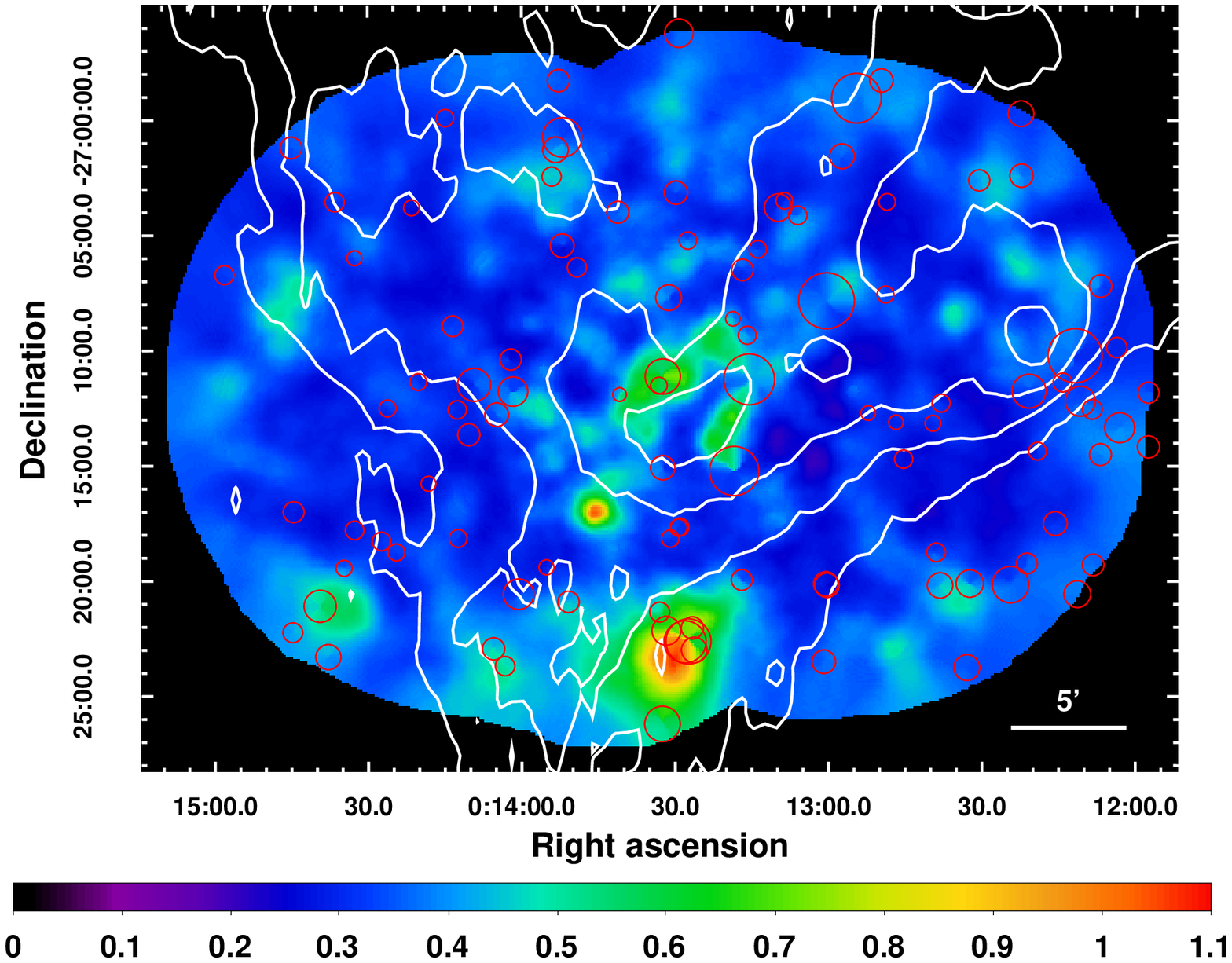}
  \includegraphics[width=0.58\linewidth]{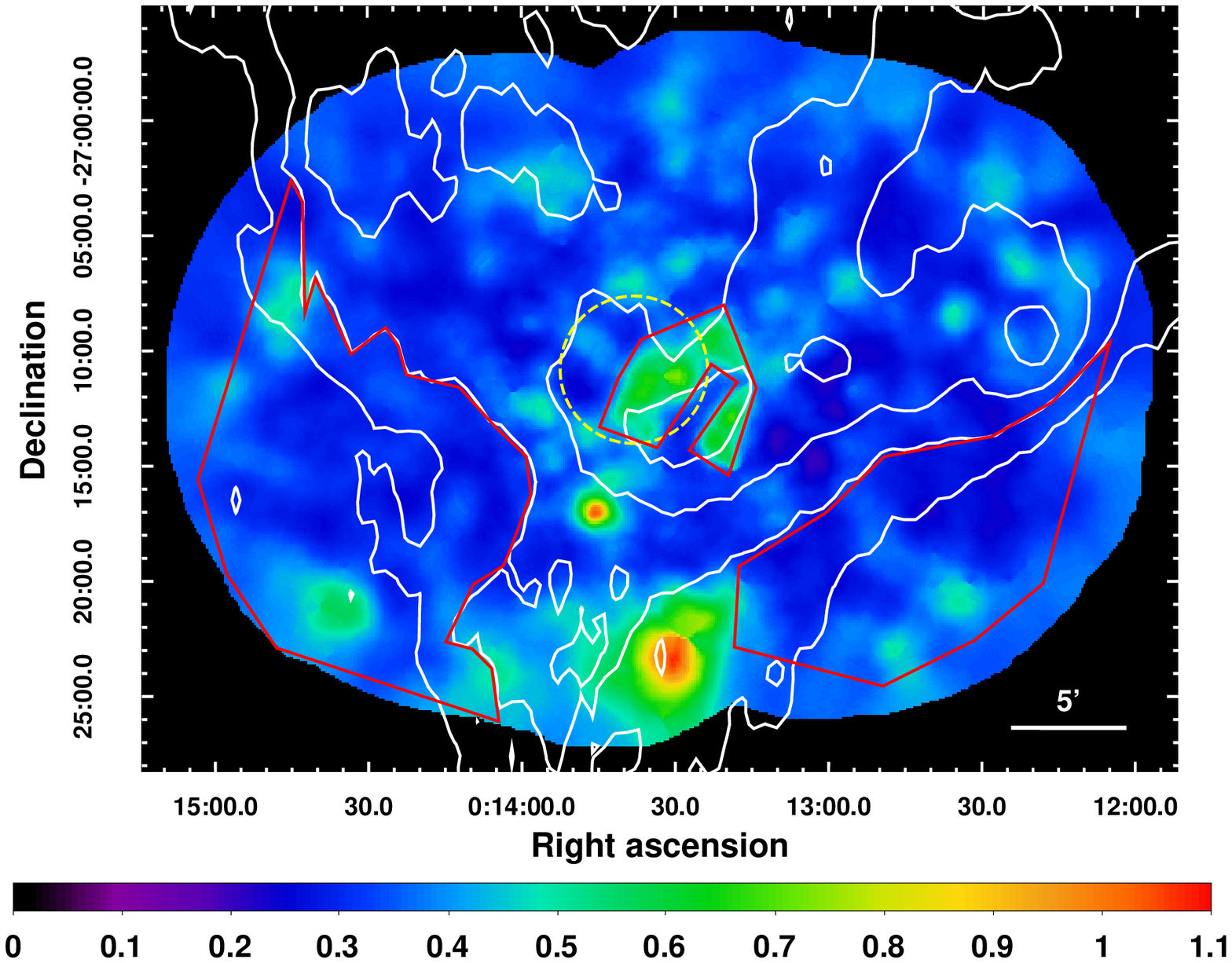}
  \caption{Mosaicked, QPB-subtracted, flat-fielded 0.4--1.2~\kev\ images of MS30.7, created by
    combining data from the MOS1, MOS2, and pn cameras. The color scales are in MOS2 counts
    \pks\ \parcminsq; the data from the other cameras were rescaled to match the response of the
    MOS2 camera. The contours indicate \HI\ column densities of 1, 2, 3, and $4 \times
    10^{20}~\pcmsq$ for a velocity interval of $-155$ to $-80~\kmps$ relative to the Local Standard
    of Rest (from a combination of Parkes \citep{bruns05} and unpublished Australia Telescope
    Compact Array (ATCA) data; C. Br{\"u}ns, 2011, private communication). In the upper panel, the
    image is overlaid with circles indicating the point source exclusion regions
    (Section~\ref{subsec:InitialDataReduction}); the holes in the data resulting from the source
    removal have been filled in using data from neighboring pixels.  In the lower panel, the image
    is overlaid with the regions from which spectra were extracted (red polygons;
    Section~\ref{subsec:SpectralExtraction}). The dashed yellow circle indicates the position and
    estimated size of the galaxy group MZ~01537 (see Section~\ref{subsec:IsEmissionFromMS30.7}).
    \label{fig:Image}}
\end{figure*}

The resulting X-ray image of MS30.7 is shown in Figure~\ref{fig:Image}. The X-ray enhancement
reported by \citet{bregman09a} is near the center of the image. The enhanced X-ray emission is
concentrated near the northern and western edges of the highest \HI\ contour. There appears to be a
gap between the northern and western portions of the enhanced emission, but this may be an artifact
due to the removal of a point source at $(\alpha,\delta) = (00^\mathrm{h}13^\mathrm{m}16^\mathrm{s},
-27\degr11\arcmin16\arcsec)$ (see upper panel of Figure~\ref{fig:Image}). Overall, the enhancement
is $\sim$6\arcmin\ across, corresponding to 100~\pc\ at a distance of 60~\kpc.

In addition to the X-ray enhancement reported by \citet{bregman09a}, there is another bright region
of similar diameter at the bottom-center of the field, as well as a smaller bright region a few
arcminutes to the south-east of the enhancement. Neither of these additional bright regions is
correlated with the HVC \HI, raising the possibility that the enhancement reported by
\citet{bregman09a} is not physically associated with MS30.7, but is just the result of a chance
alignment. If we assume that two bright regions similar in size to the enhancement is typical for a
field the size of that covered by Figure~\ref{fig:Image}, we can estimate the probability that at
least one of these bright regions will be aligned by chance with (say, within 3\arcmin\ of) the
densest part of the HVC. The field covered by Figure~\ref{fig:Image} is 1079~arcmin$^2$, and so this
probability is $1 - (1 - \pi \times 3^2 / 1079)^2 = 5\%$. This probability is not so small that we
can confidently rule out a chance alignment. However, the fact that the enhancement emission tends
to ``wrap around'' the $4 \times 10^{20}~\pcmsq$ contour does support the conclusion that the
enhancement emission is indeed physically associated with MS30.7. We will further consider this
issue in Section~\ref{subsec:IsEmissionFromMS30.7}.

\subsection{Spectral Extraction}
\label{subsec:SpectralExtraction}

For each camera from each observation, we extracted a 0.4--5.0~\kev\ spectrum of the X-ray
enhancement, and a corresponding off-enhancement spectrum, used to constrain the sky background. The
spectral extraction regions are shown in the lower panel of Figure~\ref{fig:Image}. The central
``n''-shaped region is the spectral extraction region for the enhancement -- we used the same
extraction region for both observations.  The left and right regions are the regions used to extract
the off-enhancement spectra from observations \obsB\ and \obsA, respectively. The shapes of these
regions were chosen by following the $2 \times 10^{20}~\pcmsq$ \HI\ contour and the edge of the pn
field of view.

We used the \esas\ \texttt{mos-spectra} and \texttt{pn-spectra} scripts to extract the X-ray spectra
from the data. We regrouped the X-ray spectra such that each bin contained at least 25 counts. The
spectral extraction scripts also calculated the corresponding response files needed for the analysis
-- the redistribution matrix files (RMFs) and ancillary response files (ARFs) -- using the SAS
\texttt{rmfgen} and \texttt{arfgen} programs, respectively.

From each X-ray spectrum we subtracted the corresponding QPB spectrum, calculated using the
\esas\ \texttt{mos\_back} or \texttt{pn\_back} program. Similarly to the QPB images used as part of
the X-ray image creation (Section~\ref{subsec:ImageCreation}), the QPB spectra were constructed from
a database of filter-wheel-closed data, scaled using data from the unexposed corner pixels
\citep{kuntz08a}.  The QPB spectral subtraction was carried out prior to the spectral fitting described
in the following section.

\section{SPECTRAL ANALYSIS}
\label{sec:Analysis}

\subsection{Method}
\label{subsec:Method}

We carried out our spectral analysis using
XSPEC\footnote{http://heasarc.gsfc.nasa.gov/xanadu/xspec/} version 12.7.1d \citep{arnaud96},
assuming \citet{anders89} abundances. Our basic spectral model consisted of components representing
(1) the foreground emission, (2) the Galactic halo emission, (3) the extragalactic background
emission, (4) the HVC X-ray enhancement (for the on-enhancement spectra only), and (5) components of
the instrumental background (instrumental fluorescence lines and soft proton contamination) that
were not removed by the QPB subtraction. The details of these components are as follows:

(1) We modeled the foreground emission using a single-temperature ($1T$) unabsorbed APEC model
\citep{smith01a,foster12}, the temperature of which was fixed at $kT = 0.1~\kev$ ($T = 1.2 \times
10^6~\K$). Although the foreground emission in the \xmm\ band is likely to be dominated by solar
wind charge exchange emission \citep[e.g.,][]{koutroumpa07}, such a foreground model can adequately
model the foreground emission in CCD-resolution spectra
\citep[e.g.,][]{galeazzi07,henley08a,gupta09b}. The emission measure of this component was a free
parameter -- we assumed that this emission measure was the same for the different spectral
extraction regions (see below for a justification of this assumption).

(2) We also used a $1T$ APEC model to model the diffuse Galactic halo emission. The temperature and
emission measure of this component were free parameters. We assumed that these parameters were the
same for the different spectral extraction regions (see below for a justification of the assumption
that the halo is uniform).

(3) We modeled the extragalactic background using the double broken power-law model described in
\citet{smith07a}. The normalizations of the two components were rescaled so that the
0.5--2.0~\kev\ surface brightness would match that expected from sources with fluxes below the
source removal threshold of $2 \times 10^{-15}~\flux$ ($2.99 \times 10^{-12}~\flux\ \pdegsq$, using
data from \citealt{moretti03} and \citealt{hickox06}; see Section~3.1.3 of \citealt{henley13}).

(4) We modeled the HVC X-ray enhancement with an additional $1T$ APEC model, whose temperature and
emission measure were free parameters. This component was only present in the model for the
on-enhancement spectra; in the model for the off-enhancement spectra, this component's normalization
was fixed at zero.

(5) We modeled the Al and Si instrumental fluorescence lines (at $\approx$1.49 and
$\approx$1.74~\kev, respectively) with Gaussians, whose parameters were independent for each
exposure. (Note that the pn detector does not exhibit the Si fluorescence line.)  We modeled the
soft proton contamination using a power-law not folded through the instrumental response
\citep{snowden12}.  From each exposure, we extracted two spectra: an on-enhancement spectrum and an
off-enhancement spectrum.  For each such pair of spectra, the index of the soft-proton power-law
model was the same, but the normalizations were independent. We originally tried tying together the
normalizations according to the relative scaling given by the \esas\ \texttt{proton\_scale} program,
but found that this led to poor fits above $\sim$2~\kev. The soft proton model parameters were
independent for each exposure.

The halo, enhancement, and extragalactic components were subjected to absorption, using the XSPEC
\texttt{phabs} model \citep{balucinska92,yan98}. The column density was fixed at $\NH = 1.6 \times
10^{20}~\pcmsq$, calculated from the \citet{schlegel98} $I_{100}$ maps, using the conversion
relation from \citet{snowden00}. Because we do not know exactly where the enhancement emission
arises relative to HVC material, we typically ignored absorption by the HVC itself. In order to test
the effect of ignoring absorption by the HVC, we also experimented with a variant of our model in
which we increased the column densities attenuating the enhancement and extragalactic components in
the on-enhancement spectra. Since the on-enhancement spectral extraction region lies mainly between
the $3 \times 10^{20}$ and $4 \times 10^{20}~\pcmsq$ contours (see Figure~\ref{fig:Image}), for this
model variant we increased \NH\ by $3.5 \times 10^{20}~\pcmsq$ for these two components.

We checked that the foreground and halo components of our model are indeed uniform by fitting the
above-described model (excluding the enhancement component) to the off-enhancement spectra from the
two observations, with the normalizations of the foreground and halo components independent for each
observation (although we assumed the halo temperature was the same for both observations).  The
resulting normalizations from each observation were consistent with each other. In particular, the
fact that the foreground normalizations were consistent implies that the level of solar wind charge
exchange emission is similar in the two observations.

Having confirmed the uniformity of the foreground and halo components, we fitted our full model to
our complete set of 0.4--5.0~\kev\ spectra simultaneously (a total of 12 spectra: an on-enhancement
spectrum and an off-enhancement spectrum from each of \xmm's three cameras, from each of the two
observations).

\subsection{Results}
\label{subsec:Results}

\begin{figure*}
\plottwo{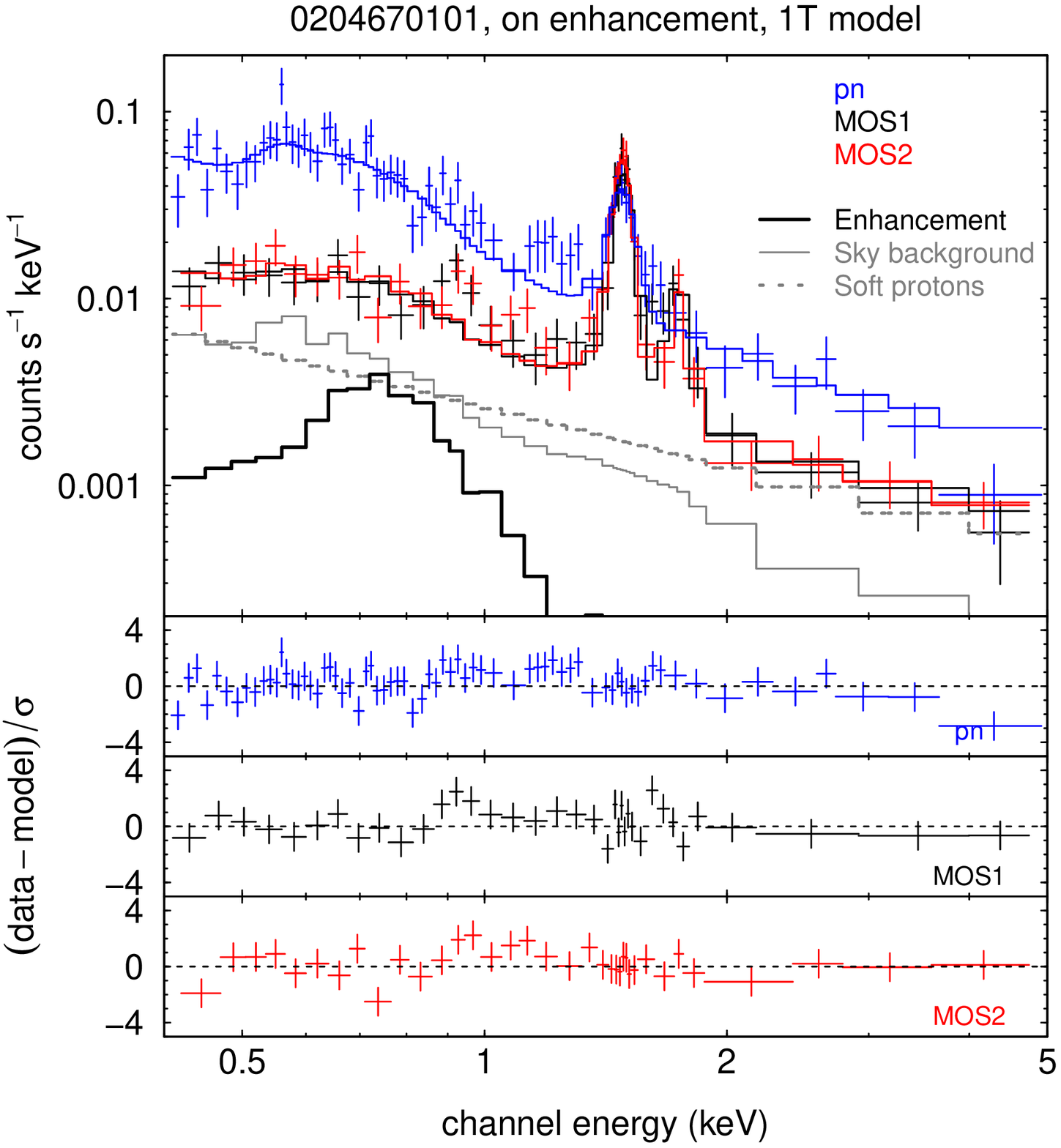}{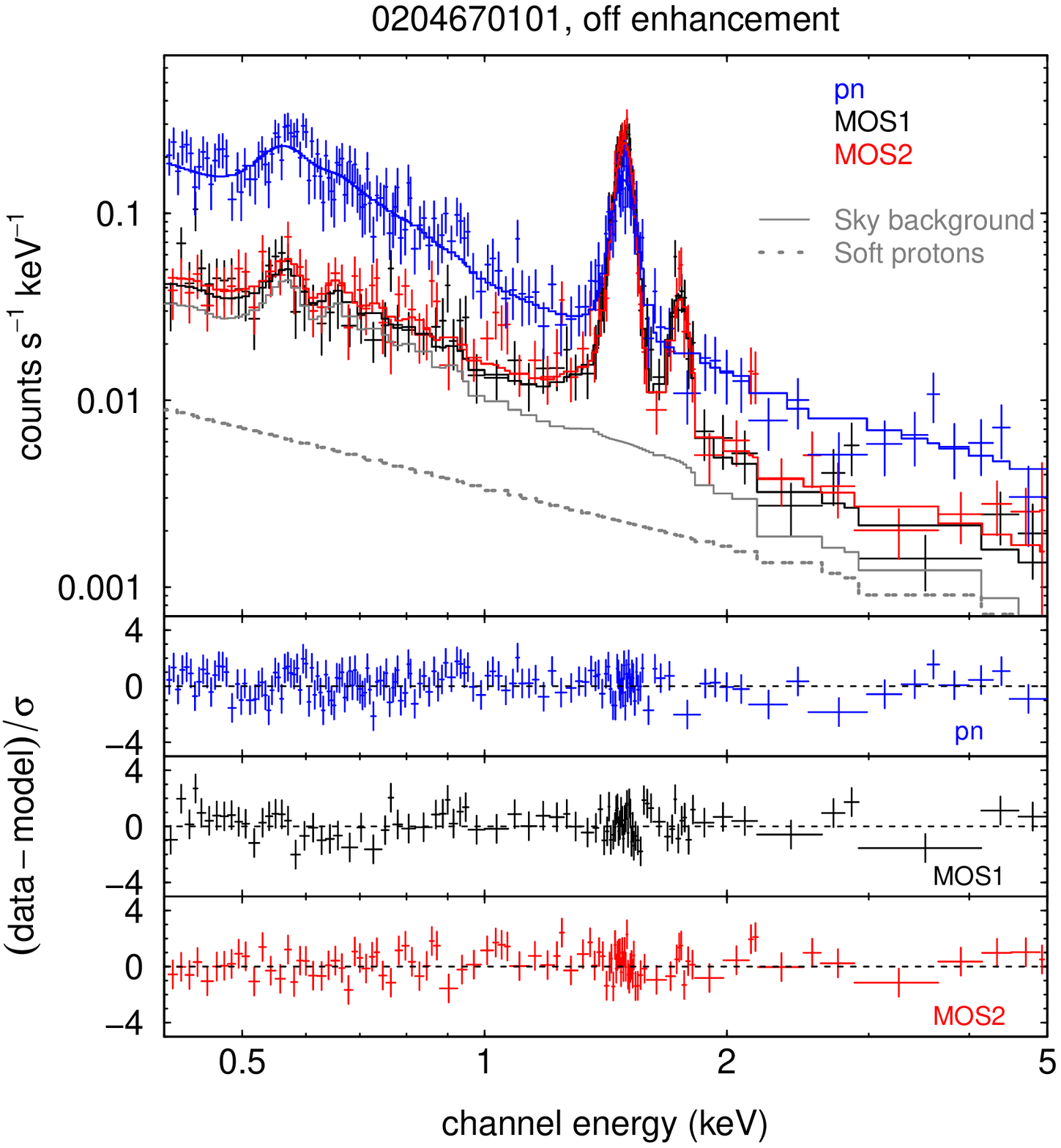}
\vspace{5mm}
\plottwo{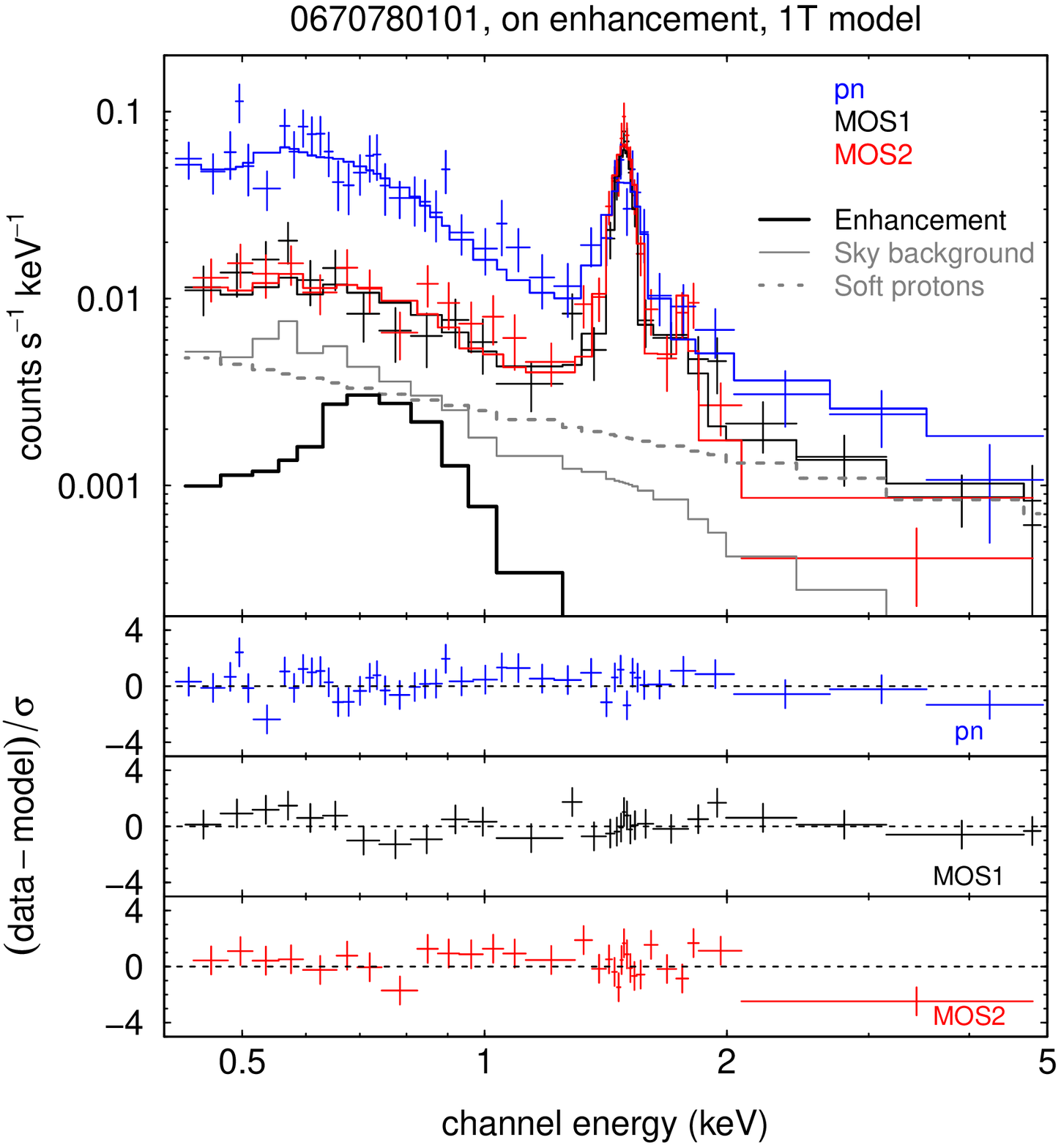}{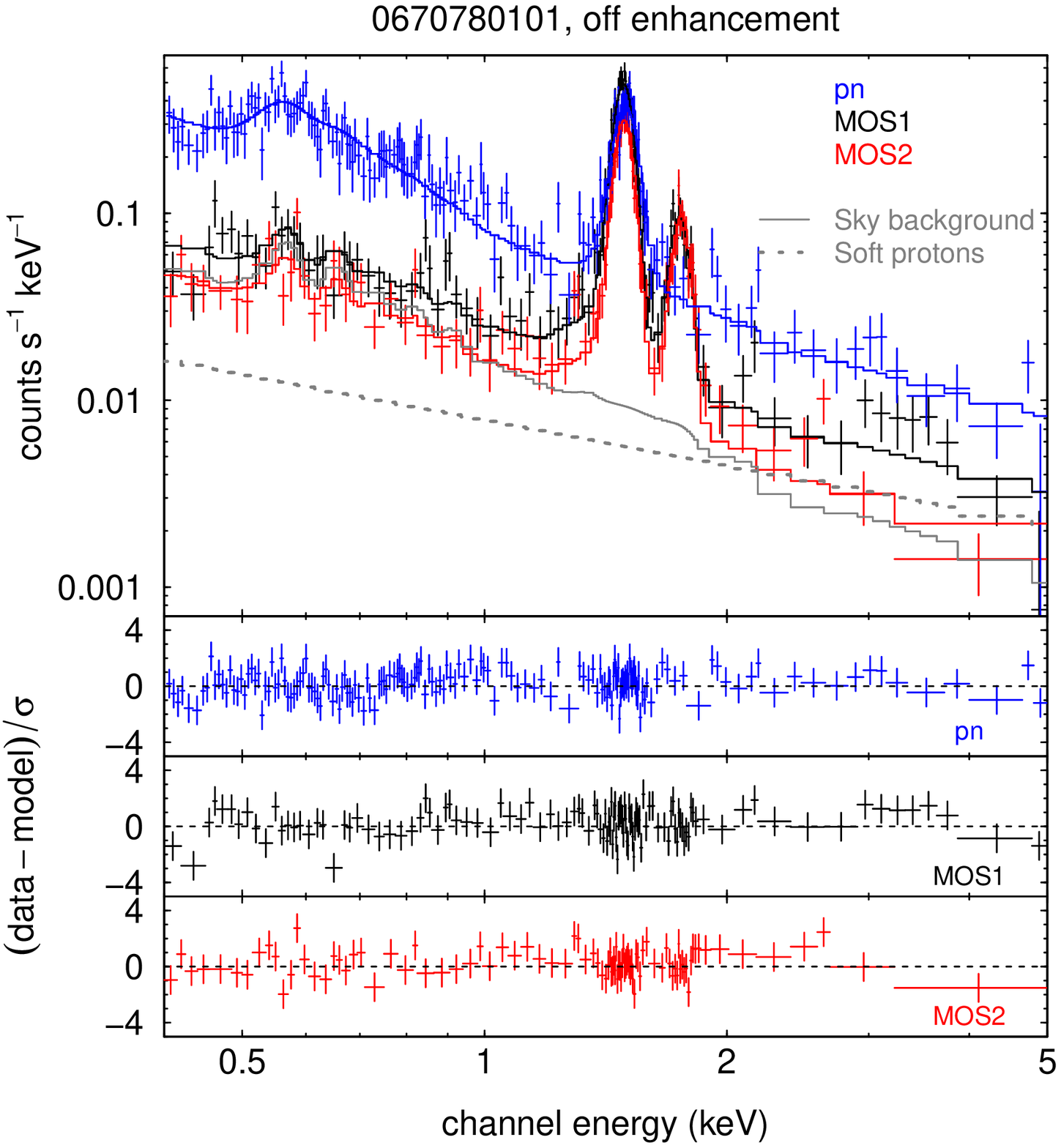}
\caption{\xmm\ spectra measured on (left) and off (right) the MS30.7 X-ray enhancement, from
  observations \obsA\ (top row) and \obsB\ (bottom row). In each plot, the main panel shows the
  X-ray spectra and best-fit models from each of the three \xmm\ cameras (see legend for color
  code). For plotting purposes only, the spectra have been grouped such that each bin has a
  signal-to-noise ratio of at least 3. Note that the count rates are higher for the off-enhancement
  spectra, as these were extracted from larger areas of the \xmm\ cameras (see lower panel of
  Figure~\ref{fig:Image}).  In addition, we show the components of the best-fit MOS1 model -- the
  enhancement itself, the sky background (foreground + halo + extragalactic), and the soft proton
  contamination are shown by a thick black line, a solid gray line, and a dashed gray line,
  respectively (to avoid clutter, we do not show the components representing the instrumental
  fluorescence lines). The three smaller panels under the main panel show the residuals for each
  camera.
  \label{fig:Spectra}}
\end{figure*}

\begin{deluxetable*}{lccccccccc}
\tablecaption{Spectral Fit Results\label{tab:Results}}
\tablehead{
                & \colhead{Foreground\tablenotemark{a}}&& \multicolumn{2}{c}{Halo}                            && \multicolumn{3}{c}{Enhancement} \\
\cline{4-5} \cline{7-9}
\colhead{Model} & \colhead{E.M.}                  && \colhead{$T$}          & \colhead{E.M.}                  && \colhead{$T$}          & \colhead{E.M.}                  & \colhead{Ne/O}      & \colhead{$\chi^2/\mathrm{dof}$} \\
                & \colhead{($10^{-3}~\emismeas$)} && \colhead{($10^6~\K$)}  & \colhead{($10^{-3}~\emismeas$)} && \colhead{($10^6~\K$)}  & \colhead{($10^{-3}~\emismeas$)} & \colhead{(solar)}
}
\startdata
Basic ($1T$)    & $9.0^{+3.9}_{-0.9}$             && $2.78^{+0.31}_{-0.09}$ & $2.98^{+0.21}_{-0.25}$          && $3.69^{+0.47}_{-0.44}$ & $2.02^{+0.43}_{-0.28}$          & 1\tablenotemark{b}  & 1824.52/1713 \\
Non-solar Ne    & $10.8^{+2.1}_{-0.8}$            && $2.93^{+0.10}_{-0.09}$ & $2.55^{+0.18}_{-0.21}$          && $3.28^{+0.23}_{-0.35}$ & $2.28^{+0.18}_{-0.40}$          & $5.5^{+2.3}_{-1.4}$ & 1787.37/1712 \\
$2T$            & $11.3^{+1.4}_{-1.3}$            && $2.95^{+0.13}_{-0.07}$ & $2.52^{+0.16}_{-0.14}$          && $2.60^{+0.36}_{-0.27}$ & $1.99^{+0.58}_{-0.39}$          & 1\tablenotemark{b}  & 1774.62/1711 \\
                &                                 &&                        &                                 && $11.9^{+1.1}_{-0.8}$   & $1.58^{+0.30}_{-0.26}$          & 1\tablenotemark{b} \\
Recombining     & $11.2^{+1.4}_{-1.0}$            && $2.95^{+0.12}_{-0.08}$ & $2.54^{+0.19}_{-0.21}$          && $3.20^{+0.52}_{-0.63}$ & \multicolumn{2}{c}{See Table~\ref{tab:Redge} for other parameters} & 1758.06/1706
\enddata
\tablecomments{Uncertainties are 90\%\ confidence intervals for a single interesting parameter.}
\tablenotetext{a}{The temperature of this component was fixed at $1.2 \times 10^6~\K$ (Section~\ref{subsec:Method}).}
\tablenotetext{b}{Frozen.}
\end{deluxetable*}

The observed spectra and the best-fit model are shown in Figure~\ref{fig:Spectra}, and the best-fit
model parameters are shown in the first row of Table~\ref{tab:Results}. The best-fit foreground
emission measure is somewhat higher than the foreground emission measures we assumed in
\citet{henley13}.  The best-fit halo temperature and emission measure are higher than the median
values obtained by \citet{henley13}, but are not outliers. Because our sky background model was
determined from spectra extracted from the same observations as our on-enhancement spectra, the fact
that the best-fit background parameters are higher than typical should not adversely affect our
measurements of the on-enhancement spectrum.

The enhancement component is hotter than the halo component, and its emission measure is similar in
magnitude to that of the halo component. The best-fit model parameters of the enhancement component
imply an intrinsic 0.4--2.0~\kev\ surface brightness of $2.6 \times 10^{-12}~\flux\ \pdegsq$. From
the size of the on-enhancement spectral extraction region (25~arcmin$^2$) and the distance of the
cloud (assumed to be 60~\kpc), we obtain the intrinsic luminosity of the enhancement: $7.9 \times
10^{33}~\ergps$.

While the fit shown in Figure~\ref{fig:Spectra} is reasonably good (reduced $\chi^2 = 1.07$), there
are some features of the spectra that are not well fit. In particular, the on-enhancement spectra
exhibit excess hard emission around 1~\kev. This excess emission is more prominent in the
obs.~\obsA\ spectra; of the obs.~\obsB\ spectra, the excess is most apparent in the MOS2
spectrum. Such excess emission is not apparent in the off-enhancement spectra, implying that the
excess hard emission originates in the X-ray enhancement.

In Section~\ref{subsec:Method}, we described a variant of our basic spectral model in which we
increased the column densities attenuating the enhancement and extragalactic components of the
on-enhancement spectra, to represent absorption by the HVC material itself. We found that this model
resulted in an enhancement temperature $0.25 \times 10^6~\K$ lower than that in
Table~\ref{tab:Results} (an insignificant difference, given the error bars), while the emission
measure and luminosity were each 40\%\ higher than those for the original model. None of these
differences is large enough to affect the discussion of physical models of the X-ray enhancement in
Section~\ref{sec:Models}.

We experimented with some additional variants of our basic spectral model. These were attempts to
improve the fit to the excess hard emission around 1~\kev\ noted above. The results of these
experiments are described in the following subsections.

\subsubsection{Non-solar Neon Abundance}

In this variant of the basic spectral model, we tried allowing the neon abundance of the enhancement
component to be a free parameter. We concentrated on neon as its strongest lines are around
1~\kev. The results are shown in the second row of Table~\ref{tab:Results}. Most of the model
parameters are not greatly affected. Thawing the neon abundance does improve the fit, but there is
still some excess on-enhancement emission just above 1~\kev\ (the enhanced neon abundance only
really has an effect at $\sim$0.9~\kev, which is the location of \NeIX\ \Kalpha). In addition, the
best-fit neon abundance is rather high: $\sim$6 times solar.

\begin{figure}
\plotone{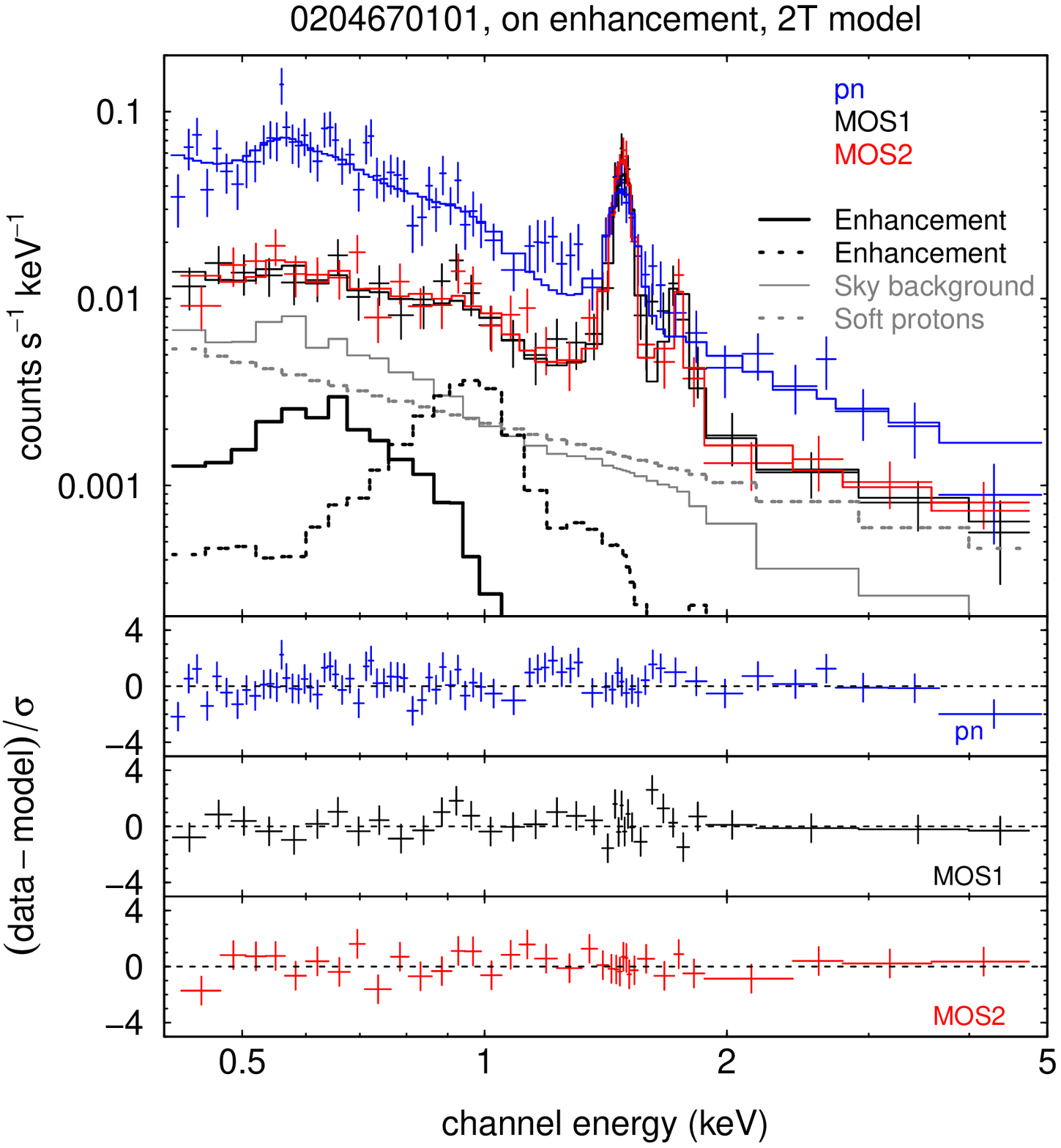}
\vspace{5mm}
\plotone{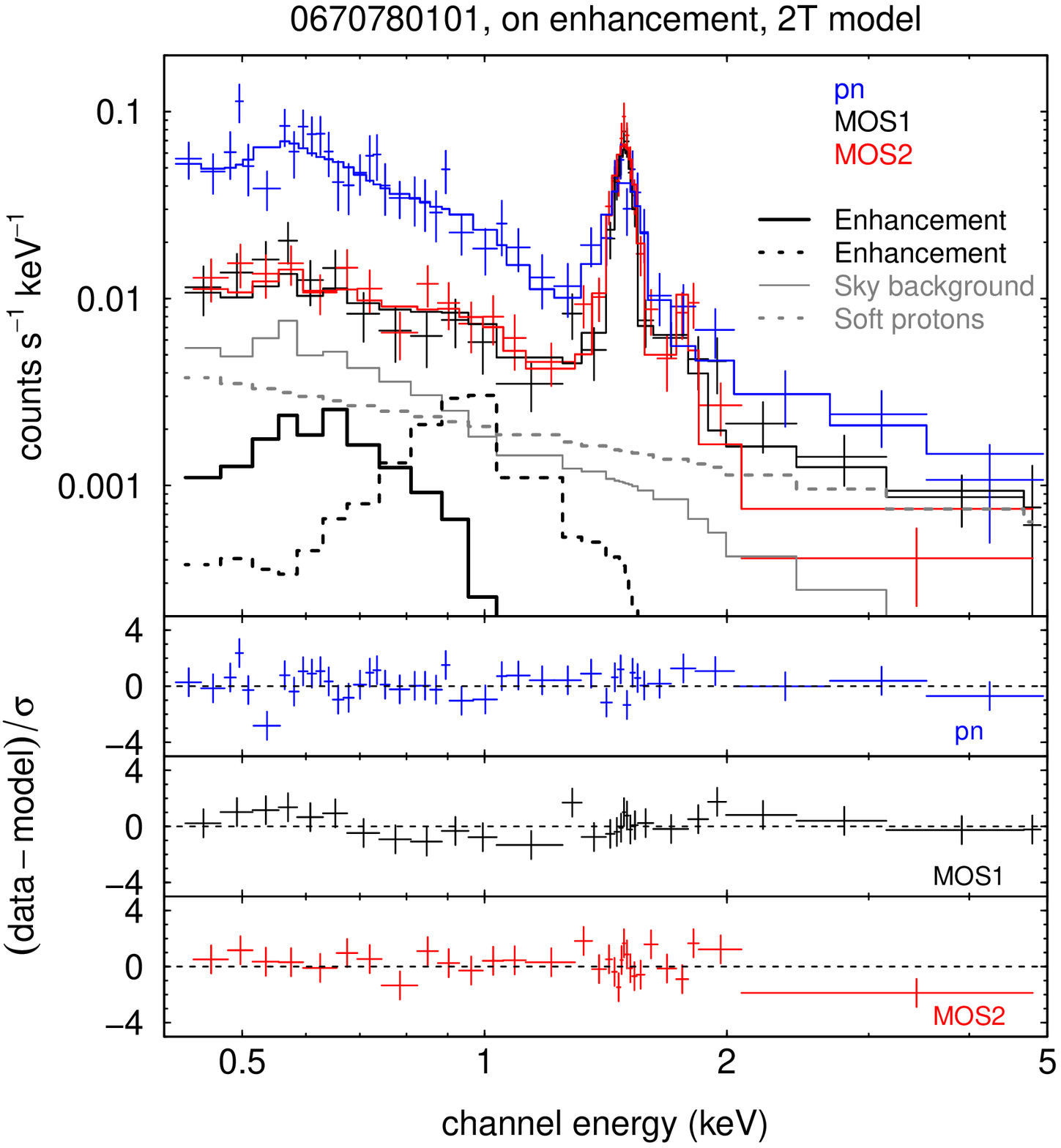}
\caption{Same as the left panels of Figure~\ref{fig:Spectra}, but using a $2T$ model to model the MS30.7
  X-ray enhancement.  The hotter enhancement component is shown by the dashed black line. The fits
  to the off-enhancement spectra (not shown) are similar to those in the right panels of
  Figure~\ref{fig:Spectra}.
  \label{fig:Spectra2T}}
\end{figure}

\subsubsection{Two-temperature Plasma Model}
\label{subsubsec:2T}

We next tried a two-temperature ($2T$) plasma model for the X-ray enhancement (as opposed to
the $1T$ model used above). The results for this model are shown in the third and fourth rows of
Table~\ref{tab:Results} (the results for the hotter enhancement component are under those for the
cooler component).  The sky background components are not greatly affected. The cooler of the two
enhancement components is somewhat cooler ($2.6 \times 10^6$ versus $3.7 \times 10^6~\K$) and
fainter than the $1T$ enhancement model. This is because the hotter component can model the harder
enhancement emission, leaving the cooler component free to shift to lower temperatures. Adding the
second enhancement component improves the fit to the on-enhancement spectra around 1~\kev\ -- see
Figure~\ref{fig:Spectra2T}. However, the best-fit temperature of this component is very high ($12
\times 10^6~\K$). The best-fit model parameters for the $2T$ model imply 0.4--2.0~\kev\ luminosities
of $6.4 \times 10^{33}$ and $7.1 \times 10^{33}~\ergps$ for the cooler and hotter components
respectively.  The total luminosity of the $2T$ model is $\sim$70\%\ larger than that of the $1T$
model.

\subsubsection{Recombining Plasma Model}
\label{subsubsec:Recombining}

In all the above spectral models, we assumed that the emission from MS30.7 is due to line emission
from a plasma in collisional ionization equilibrium (CIE). In our final variant of our basic
spectral model, we assumed that the emission is from an overionized, recombining plasma. We modeled
the enhancement as a sum of radiative recombination continua. We used the XSPEC \texttt{redge}
model, in which the flux is zero below the energy of the recombination edge, \Eedge, and varies with
photon energy $E$ as $\exp \left[ - (E - \Eedge) / kT \right]$ above the edge. Here, the temperature
$T$ is the electron temperature of the recombining plasma. We modeled the enhancement emission with
five \texttt{redge} models, representing recombinations to H-like carbon, nitrogen, and oxygen and
to He-like nitrogen and oxygen (the recombination edge for recombinations to He-like carbon is at
0.392~\kev\ \citep[Table~3.5]{dappen00}, just below the \xmm\ band that we are using). We detected
three of the five edges, and for these edges we allowed the edge energies to vary from their
expected values. For the other two edges, we fixed the edge energies at their expected values (from
\citealt{dappen00}, Table~3.5). We assumed that the electron temperature was the same for each
recombination edge.

The results for this model are shown in the final row of Table~\ref{tab:Results} (sky background
parameters and electron temperature of the recombining plasma), and in Table~\ref{tab:Redge}
(recombination edge energies and radiative recombination continuum fluxes). The electron
temperature is similar to the temperature of the $1T$ CIE enhancement model. For the three
detected edges, the measured edge energies are in good agreement with the expected values.

\begin{figure}
\plotone{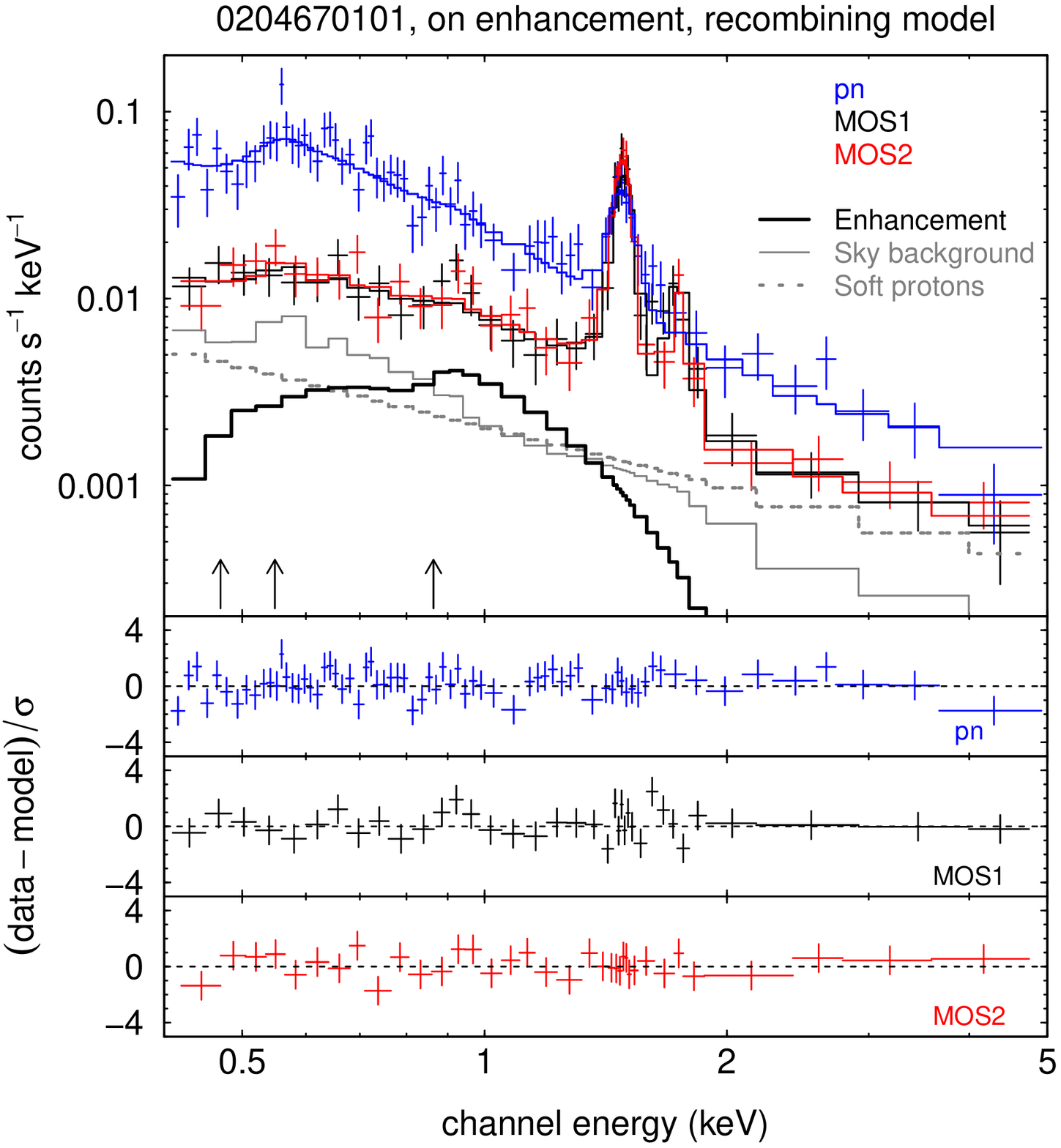}
\vspace{5mm}
\plotone{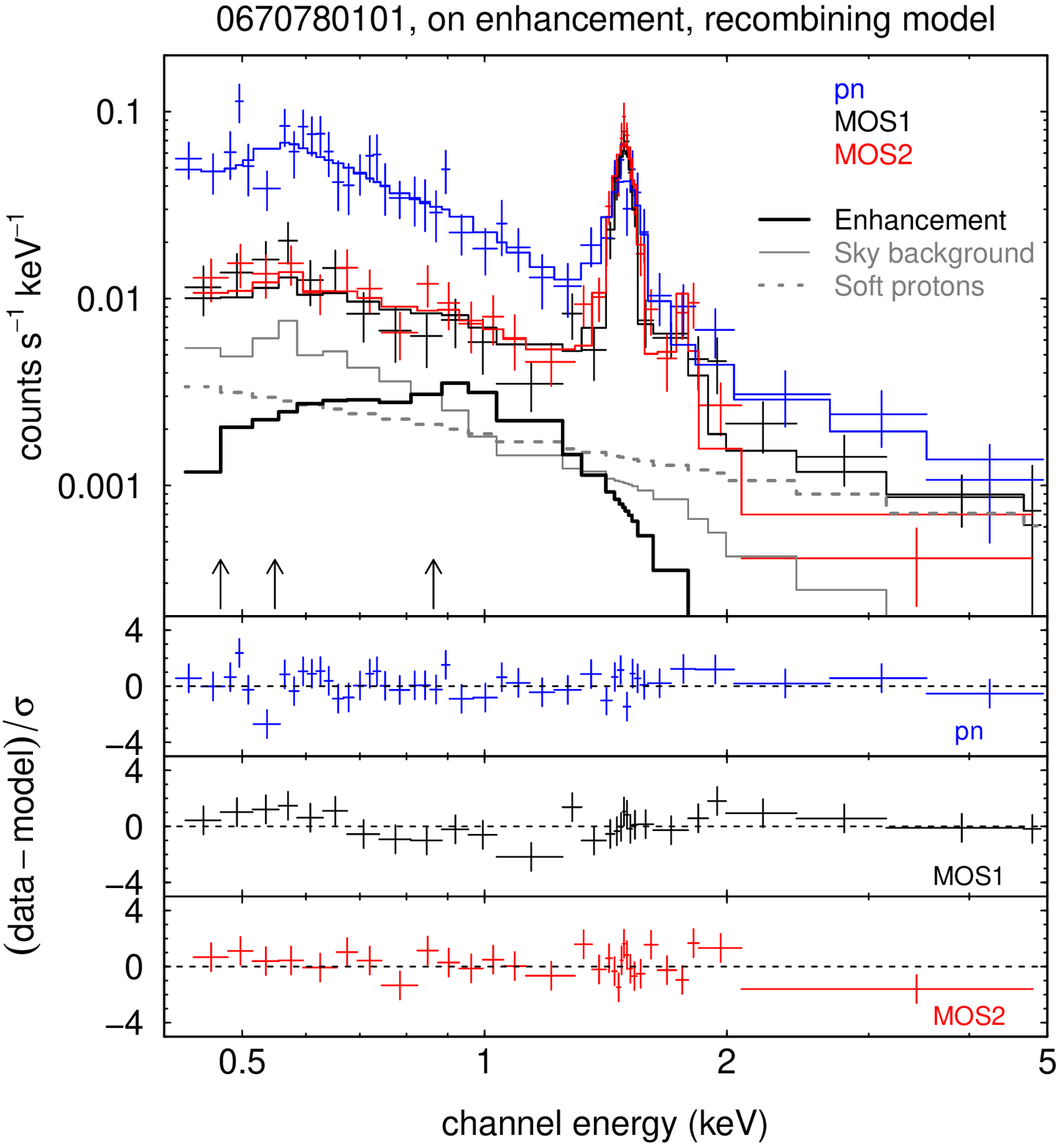}
\caption{Same as the left panels of Figure~\ref{fig:Spectra}, but using radiative recombination edges to
  model the MS30.7 X-ray enhancement (see text for details). For the enhancement model (thick black
  line), we have plotted the sum of the contributions from the individual recombination edges. The
  best-fit energies of the detected edges are indicated by the arrows (from left to right: \csix
  $\rightarrow$ \cfive, \nsix $\rightarrow$ \nfive, and \oeight $\rightarrow$ \oseven). The fits to
  the off-enhancement spectra (not shown) are similar to those in the right panels of
  Figure~\ref{fig:Spectra}.
  \label{fig:SpectraRedge}}
\end{figure}

The best-fit models for the on-enhancement spectra are shown in Figure~\ref{fig:SpectraRedge}. This
model fits the observed spectra well. In particular, as with the above-described $2T$ enhancement
model, this recombining enhancement model leads to a much better fit to the on-enhancement emission
around 1~\kev\ than our basic $1T$ enhancement model. The intrinsic 0.4-2.0~\kev\ luminosity of this
model is $1.5 \times 10^{34}~\ergps$, which is similar to that of the $2T$ enhancement model and
approximately twice that of the $1T$ model.

\begin{deluxetable*}{lccc}
\tablecaption{Fit Results for Recombining Plasma Model\label{tab:Redge}}
\tablehead{
\colhead{Recombination}         & \colhead{Expected \Eedge\tablenotemark{a}}    & \colhead{Measured \Eedge}     & \colhead{Flux} \\
                                & \colhead{(\kev)}                              & \colhead{(\kev)}              & \colhead{($10^{-7}$ photons \pcmsq\ \ps\ \parcminsq)}
}
\startdata
$\csix \rightarrow \cfive$      & 0.490                                         & $0.470^{+0.043}_{-0.049}$     & $5.1^{+7.5}_{-1.2}$ \\
$\nsix \rightarrow \nfive$      & 0.552                                         & $0.549^{+0.048}_{-0.072}$     & $4.2^{+1.2}_{-1.1}$ \\
$\nseven \rightarrow \nsix$     & 0.667                                         & 0.667\tablenotemark{b}        & $<$0.67 \\
$\oseven \rightarrow \osix$     & 0.739                                         & 0.739\tablenotemark{b}        & $<$0.48 \\
$\oeight \rightarrow \oseven$   & 0.871                                         & $0.864^{+0.044}_{-0.050}$     & $1.60^{+0.51}_{-0.71}$
\enddata
\tablecomments{Uncertainties are 90\%\ confidence intervals for a single interesting parameter.
  See the final row of Table~\ref{tab:Results} for the parameters of the sky
  background model and the electron temperature of the recombining plasma.}
\tablenotetext{a}{\citet{dappen00}, Table~3.5.}
\tablenotetext{b}{Frozen.}
\end{deluxetable*}

For nitrogen and oxygen, we can use the ratios of the measured recombination fluxes to infer the
$\nsix/\nseven$ and $\oseven/\oeight$ ion ratios, and hence the ionization temperatures of these two
elements. If $F_X$ is the flux due to recombinations from the $+X$ ion to the $+(X-1)$ ion, then
$F_X \propto \Ne n_X R_X$, where \Ne\ is the electron number density, $n_X$ is the number density of
the $+X$ ion, and $R_X$ is the $+X \rightarrow +(X-1)$ radiative recombination coefficient. Hence,
\begin{equation}
  \frac{n_{X-1}}{n_X} = \frac{F_{X-1}}{F_X} \frac{R_X}{R_{X-1}}.
\end{equation}

For oxygen, taking the upper limit of the $\oseven \rightarrow \osix$ recombination flux and the
lower limit of the $\oeight \rightarrow \oseven$ flux, we find $F_7 / F_8 < 0.54$. The ratio of
recombination coefficients is $R_8 / R_7 = 1.69$ (\citealt{verner96}; we evaluated this ratio at $T
= 3.2 \times 10^6~\K$, the best-fit electron temperature for the recombining plasma, but in fact
this ratio varies very slowly with temperature). Hence, $n_7 / n_8 < 0.91$. This corresponds to an
ionization temperature for oxygen of $>$$2.8 \times 10^6~\K$ (using ionization balance data from
\citealt{mazzotta98}).

Repeating the above analysis for nitrogen, we find $F_6 / F_7 > 4.67$ (note that the ``7'' subscript
now refers to \nseven\ rather than \oseven), $R_7 / R_6 = 1.78$ (\citealt{verner96}; again, this
ratio varies very slowly with temperature), and hence $n_6 / n_7 > 8.3$, corresponding to an
ionization temperature for nitrogen of $<$$1.3 \times 10^6~\K$ \citep{mazzotta98}. Note that the
inferred ionization temperature for nitrogen is much lower than that for oxygen. Note also that the
nitrogen ionization temperature is less than the electron temperature -- in this situation, one
would expect the nitrogen to be ionizing rather than recombining.

The fact that the nitrogen ionization temperature is less than the electron temperature suggests
that our spectral model, which is characterized by a single electron temperature for the entire
recombining plasma, may be overly simplistic. Unfortunately, our spectral data are insufficient to
conclude whether or not recombination emission really is a major contributor to the observed
emission -- all we can do is note the improvement to the fit around 1~\kev. However, because this
model implies an oxygen ionization temperature that in turn implies a high shock speed (see
Section~\ref{subsec:ShockHeating}), and because the luminosity of this model is similar (within a
factor of 2) to those of the $1T$ and $2T$ enhancement models, the conclusions of the following
Section, in which we discuss physical models of the X-ray enhancement, are the same whether we
assume this model or a CIE model is the best description of the observed emission.

\section{MODELS OF THE X-RAY ENHANCEMENT}
\label{sec:Models}

In this section, we consider different physical models for the X-ray enhancement. In
Section~\ref{subsec:TurbulentMixing}, we consider the possibilities that the emission is due to
turbulent mixing of cold cloud material with a hot ambient medium, or to compression of such a
medium by the cloud. We then examine the possibility that the observed emission is from a
shock-heated plasma (Section~\ref{subsec:ShockHeating}). In Section~\ref{subsec:ChargeExchange}, we
consider charge exchange reactions between the cold cloud and a hot ambient medium as a possible
source of the emission. Finally, in Section~\ref{subsec:MagneticReconnection}, we return to the idea
that the X-ray emission is from a hot plasma, but heated by magnetic reconnection rather than by a
shock. However, before examining the individual models, we will first discuss the likely physical
parameters for such models.

\subsection{Model Parameters}
\label{subsec:ModelParameters}

The important physical parameters for the following models are the density, temperature, and magnetic
field of the ambient medium, and the radius and speed of the cloud.

The ambient conditions in the vicinity of the Magellanic Stream are not well known. X-ray
observations imply halo temperatures of $\sim$$(\mbox{1--3}) \times 10^6~\K$
\citep[e.g.,][]{kuntz00,yoshino09,henley13,gupta13}.  It is uncertain out to what distance such
temperature measurements are applicable. However, the conclusions reached below are not very
sensitive to the ambient temperature.

For the density of a hot ambient medium, we will typically assume a value of
$10^{-4}~\pcc$. \citet{bregman09a} point out that if the halo density were more than a few times
$10^{-4}~\pcc$, the dispersion measure would exceed the values measured toward some Large Magellanic
Cloud (LMC) pulsars. For example, if the hot halo density were $5 \times 10^{-4}~\pcc$ in the
vicinity of MS30.7 (and presumably higher than this closer to the Milky Way), the dispersion measure
toward the LMC ($d = 50~\kpc$) would be $\ge$$72~\dm$: $\ge$$25~\dm$ from the hot halo, plus
47~\dm\ from the warm ionized medium (using the best-fit model of \citealt{gaensler08}, and taking
into account the LMC's Galactic latitude). In contrast, the two lowest dispersion measures measured
toward the LMC are 45 and 65.8~\dm\ (there is some uncertainty as to whether the pulsar that yields
the lower value is in the LMC or in the foreground; \citealt{manchester06}).  Furthermore, if the
hot halo density were $5 \times 10^{-4}~\pcc$ out to a distance of at least 60~\kpc\ (the assumed
distance of MS30.7), the hot halo's emission measure would exceed 0.015~\emismeas. In contrast,
analyses of the soft X-ray background emission typically yield halo emission measures of a few times
$10^{-3}~\emismeas$ (e.g., \citealt{kuntz00,yoshino09,henley13}; note that none of these studies
report an emission measure exceeding 0.01~\emismeas).

In Section~\ref{subsec:ShockHeating}, we will consider the possibility that MS30.7 is ramming into
material shed from a preceding cloud in the Stream. In this situation, the ambient material will
have a higher density and a lower temperature than that discussed above. Note that, in this case,
the higher ambient density will be relatively localized, and so will not violate the above density
constraints.

When we discuss magnetic reconnection (Section~\ref{subsec:MagneticReconnection}), we will need to
know the ambient magnetic field. There are no direct measurements of the ambient field strength in
the vicinity of the Magellanic Stream. \citet{mccluregriffiths10} used extragalactic rotation
measures to estimate the magnetic field in an HVC in the Leading Arm of the Magellanic System. They
found that the line-of-sight component of the coherent magnetic field was
$\ga$6~\microgauss. However, it should be noted that this value pertains to the field within the
cloud, which may be enhanced relative to the ambient field. Furthermore, this HVC is only
$\sim$10~\kpc\ above the Galactic disk (using the distance assumed by \citealt{mccluregriffiths10}),
compared with $\sim$60~\kpc\ for MS30.7. If we assume equipartition between the ambient thermal and
magnetic energy densities, the ambient magnetic field strength is
\begin{align}
  B_\mathrm{equip} &= \sqrt{12 \pi nkT} \nonumber \\
                 &= (1.0~\microgauss)
                    \left( \frac{\phantom{X}n\phantom{X}}{10^{-4}~\pcc} \right)^{1/2}
                    \left( \frac{T}{2 \times 10^6~\K} \right)^{1/2},
\end{align}
where $n$ and $T$ are the ambient density and temperature, respectively.

For the cloud itself, we will assume a radius $r \approx 50~\pc$ (from the size of the X-ray
enhancement; Section~\ref{subsec:ImageCreation}). For the cloud speed, we will assume that the
orbital speed of the Magellanic Stream is similar to those of the Magellanic
Clouds. \citet{kallivayalil06a,kallivayalil06b} used the \textit{Hubble Space Telescope} to measure
the proper motions of the Magellanic Clouds. Combining these measurements with radial velocities
measurements, they obtained speeds relative to the Galactic Center of $378 \pm 18$ and $302 \pm
52~\kmps$ for the LMC and the Small Magellanic Cloud (SMC), respectively. Here, we will follow
\citet{blandhawthorn07}, and assume an orbital speed of 350~\kmps\ for the Magellanic Stream.  Given
the orbital speeds of the Magellanic Clouds, MS30.7's speed is unlikely to exceed $\sim$400~\kmps.

\subsection{Turbulent Mixing with or Compression of a Hot Ambient Medium}
\label{subsec:TurbulentMixing}

The X-ray enhancement could in principle arise from turbulent mixing of cool HVC material with a hot
ambient medium, resulting in gas of intermediate density and temperature that is potentially
brighter in X-rays than the background \citep{shelton12}.  However, hydrodynamical simulations of
HVCs traveling through hot (typically $1 \times 10^6~\K$) ambient gas imply that, in practice, only
slight enhancements in the X-ray emission result from turbulent mixing \citep[specifically, their
  Case~A models]{shelton12}. We re-examined the model spectra created by \citeauthor{shelton12}\ from
their A models, calculating the 0.4--2.0~\kev\ surface brightnesses for comparison with the measured
value for MS30.7. We ignored models A8--A10, as the clouds in these models were initialized with
supersonic speeds, and so shock heating (discussed below) would tend to mask the effects of
turbulent mixing. Of the remaining A models, we found that the brightest had a peak
0.4--2.0~\kev\ surface brightness of $2.6 \times 10^{-14}$~\flux\ \pdegsq\ (model A1, at $t =
30~\Myr$; at this time, the ambient density in the vicinity of the cloud is a few times
$10^{-4}~\pcc$). This is two orders of magnitude less than the intrinsic surface brightness of the
MS30.7 X-ray enhancement (Section~\ref{subsec:Results}).

The X-ray enhancement could also result from the compression of a hot ambient medium by the cloud,
where the increased density results in an increase in the X-ray brightness
\citep{herbstmeier95}. \citet{shelton12} did not consider this mechanism in their study, but the
faintness of their A models also rules this out as the mechanism responsible for the MS30.7
enhancement.

\subsection{Shock Heating}
\label{subsec:ShockHeating}

\subsubsection{Strong Shocks}
\label{subsubsec:StrongShocks}

If we assume that the X-ray-emitting plasma is due to shock heating, we can translate the measured
temperature to a corresponding shock speed. First, let us consider a strong (i.e., high Mach number)
shock. Such a strong shock could arise in the context of MS30.7 if the cloud were ramming into cool
material shed from a preceding cloud. For material crossing a strong shock at speed $v$, the
postshock temperature is \citep[e.g.,][]{dyson97}
\begin{align}
  T &= \frac{3 \bar{m} v^2}{16k} \nonumber \\
    &= (1.4 \times 10^5~\K) \left( \frac{v}{100~\kmps} \right)^2,
  \label{eq:Postshock}
\end{align}
where $\bar{m} \approx 1 \times 10^{-24}~\gram$ is the average mass per particle. Note that the
speed in the above expression is the speed at which material crosses the shock, which may be
somewhat faster than the speed of the cloud, as the shock tends to move away from the cloud as the
cloud and shock evolve, at least in the early stages of the cloud's evolution.  Hydrodynamical
simulations imply that the shock speed exceeds the cloud speed by $\la$10\%\ for strong shocks in
cool and warm ambient media induced by initially round clouds \citep{shelton12}. Note also that, in
practice, the average post-shock temperature behind an HVC's bow shock would be lower than that
expected from Equation~(\ref{eq:Postshock}) for a number of reasons, especially if the cloud is
traveling through relatively dense cool or warm gas: (1) gas toward the side of the cloud will hit
the bow shock obliquely, reducing the postshock temperature, (2) the cloud will decelerate as it
passes through the dense gas, weakening the shock, and (3) radiative cooling will be important in
the dense shocked gas.

For the $1T$ enhancement model, the best-fit temperature of the enhancement
(Table~\ref{tab:Results}) implies a shock speed of 510~\kmps. For the $2T$ enhancement model, the
two components' best-fit temperatures correspond to shock speeds of 430 and 920~\kmps,
respectively. These speeds are unreasonably high for MS30.7, even allowing for the fact that the
shock speed may be greater than the cloud speed (Section~\ref{subsec:ModelParameters}). A shock
speed of 385~\kmps\ (10\%\ greater than the assumed orbital speed of the Magellanic Stream) would
yield a postshock temperature that is about half the observed $1T$ value.

\subsubsection{Shock Heating of a Hot Ambient Medium}
\label{subsubsec:ShockHeatingHotMedium}

If, instead of ramming into cool material, MS30.7 is traveling through a hot ($\sim$$10^6~\K$)
ambient medium, the Mach number will be lower than in the strong-shock case considered above. In
this case, we can use the general formula relating the pre- and postshock temperatures, $T_1$ and
$T_2$, respectively, to the preshock Mach number, $M_1$ \citep[Equation~(15.37)]{shu92}:
\begin{equation}
  \frac{T_2}{T_1} = \frac{ \left[ (\gamma+1) + 2\gamma(M_1^2-1) \right] \left[ (\gamma+1) + (\gamma-1)(M_1^2-1) \right] }{(\gamma+1)^2 M_1^2}.
  \label{eq:Postshock2}
\end{equation}
For an ambient temperature of $T_1 = 1 \times 10^6~\K$ and for $T_2$ equal to the best-fit
temperature of the $1T$ enhancement model (Table~\ref{tab:Results}), the above equation implies $M_1
= 3.0$, or a shock speed of 460~\kmps\ (the speed of sound in $1 \times 10^6~\K$ gas is
$\approx$150~\kmps). The best-fit temperatures of the $2T$ enhancement model imply shock speeds of
360 and 900~\kmps, respectively.

As discussed in Section~\ref{subsubsec:StrongShocks}, the shock in front of an HVC may travel faster
than the cloud itself. For a hot, low-density ambient medium, we find that the difference in speeds
is greater than in a denser, cooler medium. We carried out one-dimensional simulations of shock
tubes in which gas initially traveling at 400~\kmps\ (representing HVC material) rams into
stationary low-density hot gas (hydrogen number density $\nH = 6.45 \times 10^{-5}~\pcc$, $T =
10^6~\K$) or denser warm gas ($\nH = 6.45 \times 10^{-3}~\pcc$, $T = 10^4~\K$). We found that the
shocks propagated into the hot and warm stationary gas at $\approx$570 and $\approx$405~\kmps,
respectively. Hence, a cloud speed of $\sim$350~\kmps\ (Section~\ref{subsec:ModelParameters}) could
plausibly produce a post-shock temperature similar to that of our best-fit $1T$ enhancement model.

While a shock in a hot ambient medium could approximately reproduce the observed $1T$ temperature of
the enhancement, the low density of the shocked material will result in emission much too faint to
explain the observations. The shock speeds estimated above using Equation~(\ref{eq:Postshock2})
imply shock compression ratios of $\sim$3 \citep[Equation~(15.35)]{shu92}. If the ambient electron
density in the vicinity of the Magellanic Stream is $\Ne \sim 10^{-4}~\pcc$
(Section~\ref{subsec:ModelParameters}), then behind the shock $\Ne^2 \sim 10^{-7}~\cm^{-6}$. Since
MS30.7 is traveling close to perpendicular to the line of sight,\footnote{From the assumed orbital
  speed (350~\kmps; Section~\ref{subsec:ModelParameters}) and the line-of-sight velocity (118~\kmps;
  \citealt{bregman09a}) of MS30.7, the angle between MS30.7's velocity vector and the line of sight
  is 70\degr.} the extent of the X-ray enhancement along the line of sight is likely similar to its
extent on the sky (i.e., $\sim$100~\pc; Section~\ref{subsec:ImageCreation}). This implies an
emission measure of $\sim$$10^{-5}~\emismeas$, two orders of magnitude less than the observed
emission measures in Table~\ref{tab:Results}.

The postshock emission measure could be increased by increasing the ambient density, but this
density is unlikely to substantially exceed $10^{-4}~\pcc$ at the distance of the Magellanic Stream
(Section~\ref{subsec:ModelParameters}). If we take $5 \times 10^{-4}~\pcc$ as an upper limit on the
ambient density in the vicinity of the Magellanic Stream, the postshock emission measure will be
$\sim$$2 \times 10^{-4}~\emismeas$, which is an order of magnitude smaller than the observed
value. Our conclusion here, that shock heating of a hot ambient medium would result in emission that
is too faint to explain the MS30.7 observations, is consistent with that in
Section~\ref{subsec:TurbulentMixing}, where we stated that \citepossessive{shelton12} Case~A models
implied that compression of a hot ambient medium would result in emission that is too faint
(although in that case the compression was not necessarily via a shock).

\subsubsection{Predictions from \citet{shelton12} HVC Models}
\label{subsubsec:SheltonModels}

We further investigated shock heating using the Case~B hydrodynamical models of \citet{shelton12}.
In these models, the initially spherical cloud hits warm gas (possibly representing material shed
from a preceding cloud; $T = 10^4~\K$, $\nH = 6.45 \times 10^{-3}~\pcc$), after passing through hot
halo gas ($T = 10^6~\K$, $\nH = 6.45 \times 10^{-5}~\pcc$). Note that the temperatures and densities
of the ambient gases match those of the stationary gases in the shock tube simulations described in
Section~\ref{subsubsec:ShockHeatingHotMedium}.  \citeauthor{shelton12}\ ran models with and without
radiative cooling enabled (Br and Ba, respectively). The model clouds had a range of initial speeds
(200--400 and 200--600~\kmps\ for the Br and Ba models, respectively). The number in each model's
name (e.g., Ba3) indicates the model cloud's initial speed in units of 100~\kmps. The Ba models were
run for up to 28~\Myr, with data output every 2~\Myr. The Br models were run for 2~\Myr, with data
output every 40~kyr. (Br model data are also available at 2~\Myr\ intervals beyond
$t=2~\Myr$. However, we found that these models were too soft and faint to explain the observations,
and we do not show the results for these later Br epochs below.)

For each epoch of each model, we calculated the spectrum averaged over a radius of 50~\pc\ from the
cloud center (for an observer looking directly along the cloud's velocity vector). The spectra were
calculated assuming CIE (see \citealt{shelton12} for more details of the spectral calculations).  We
did not subtract off the contribution from the ambient medium in the model, to avoid potentially
having to deal with negative model count rates. However, the model ambient medium is over 1000 times
fainter than observed X-ray enhancement, so if an HVC model well matches the observed brightness of
the enhancement, the contribution from the model ambient medium will be negligible.

We compared the surface brightnesses of the above-calculated spectra with the intrinsic surface
brightness of the X-ray enhancement inferred from the spectral fitting
(Section~\ref{subsec:Results}). In addition, we used each model spectrum as the enhancement
component of our spectral model (Section~\ref{subsec:Method}), with the normalization as a free
parameter. From these fits, we obtained $\chi^2$ as a function of model epoch for each model that we
investigated, allowing us to see how well the predicted spectra match the shape of the observed
enhancement spectrum.

\begin{figure*}
  \centering
  \plottwo{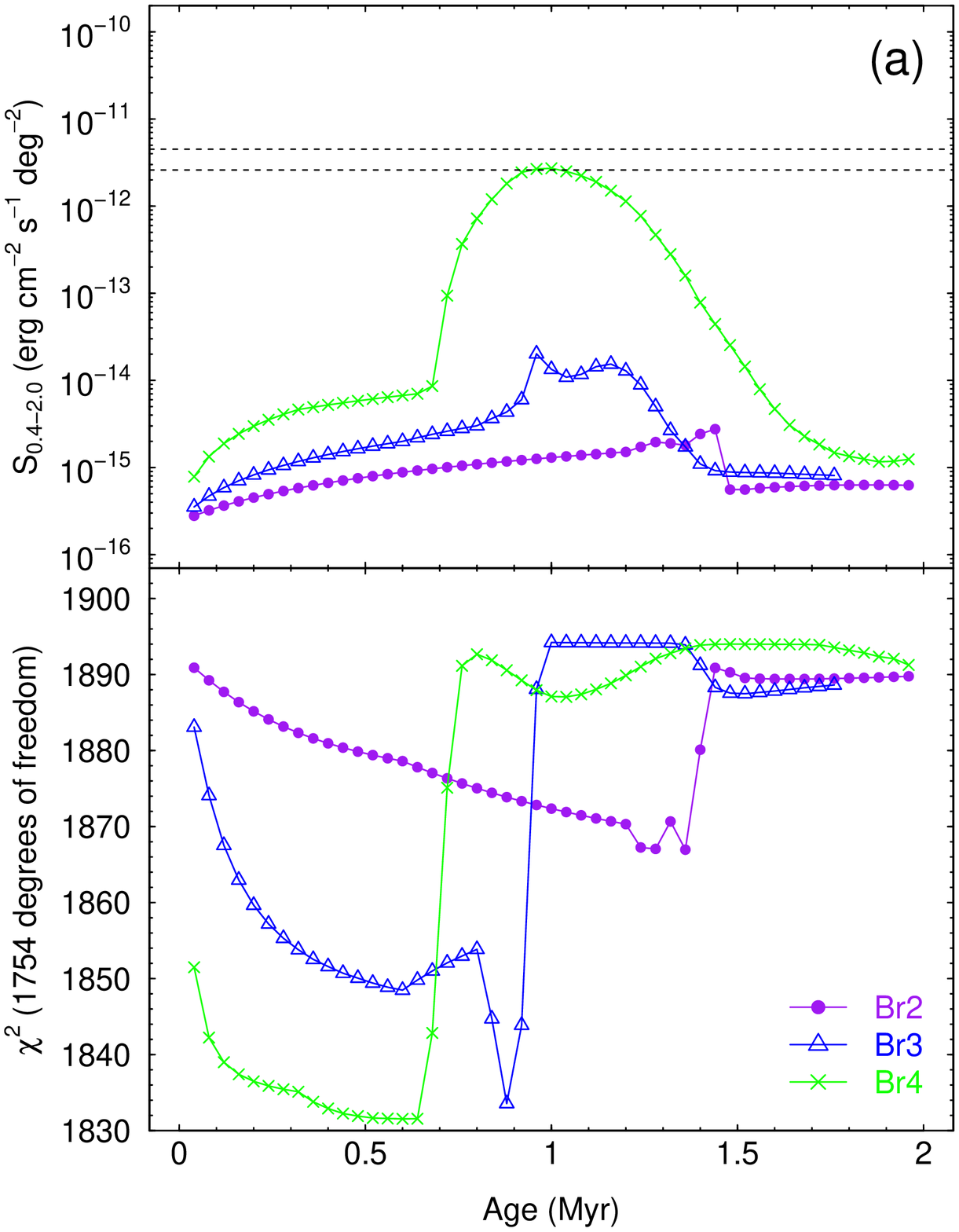}{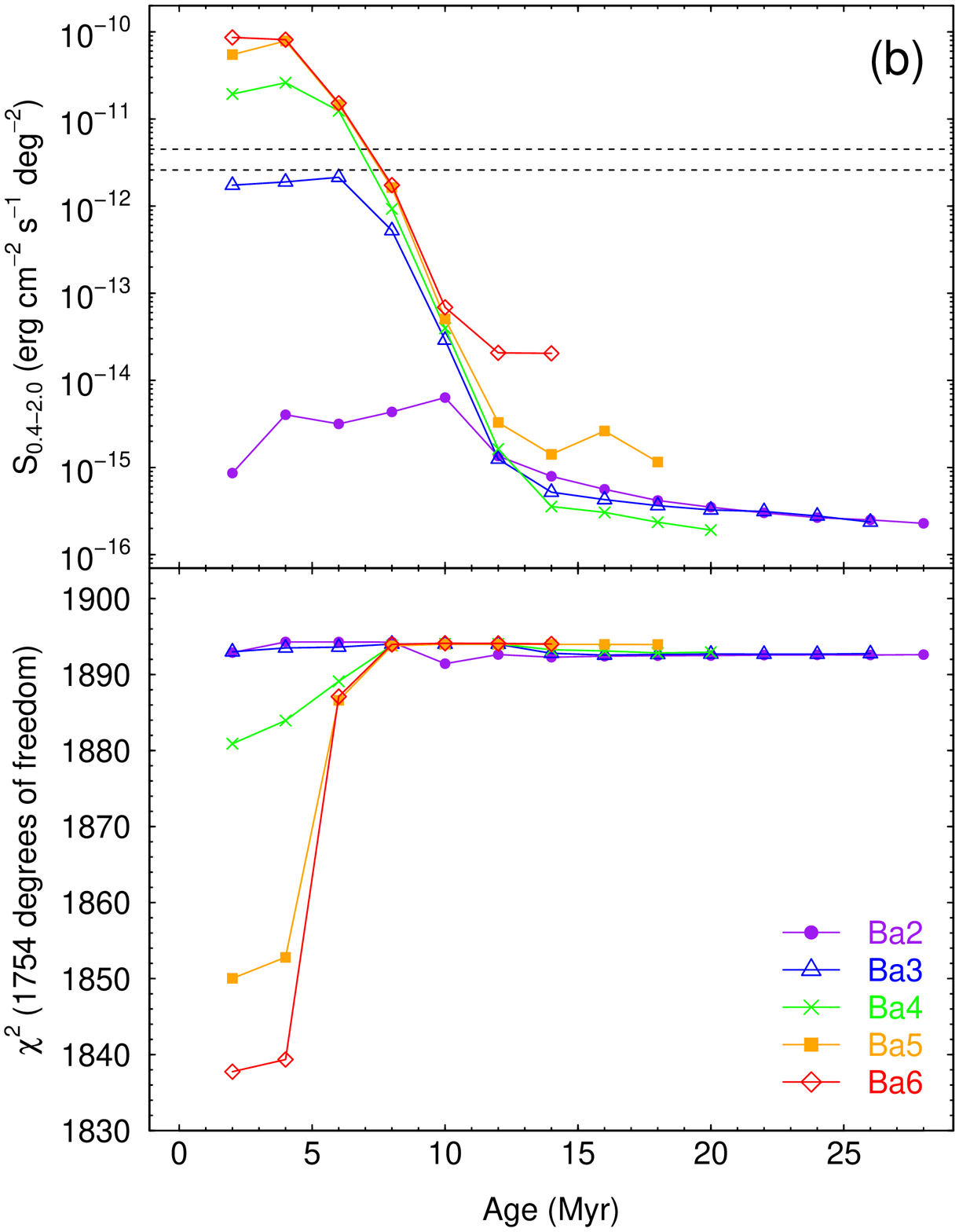}
  \caption{Intrinsic 0.4--2.0~\kev\ surface brightnesses, $S_{0.4-2.0}$, of various HVC models from
    \citet{shelton12} (top) and $\chi^2$ from fits in which the MS30.7 X-ray enhancement is modeled
    using said HVC models (bottom), as functions of model epoch.  Plot (a) shows the results for the
    Br models, in which radiative cooling was enabled, while plot (b) shows the results for the Ba
    models, in which radiative cooling was not enabled (note the different ranges on the time
    axes). The number in each model name indicates the model cloud's initial speed in units of
    100~\kmps. The horizontal dashed lines indicate the intrinsic surface brightnesses of the MS30.7
    enhancement inferred from the $1T$ (lower line) and $2T$ (upper line) models
    (Section~\ref{subsec:Results}).  Note that the number of degrees of freedom (1754) is larger
    than the numbers of degrees of freedom in Table~\ref{tab:Results}, because here we fixed
    instrumental lines' energies and widths at the best-fit values obtained when fitting the $1T$
    enhancement model (otherwise we found that XSPEC ran into problems during some of the fits).
    \label{fig:HVCresults}}
\end{figure*}

\begin{figure*}
  \centering
  \plottwo{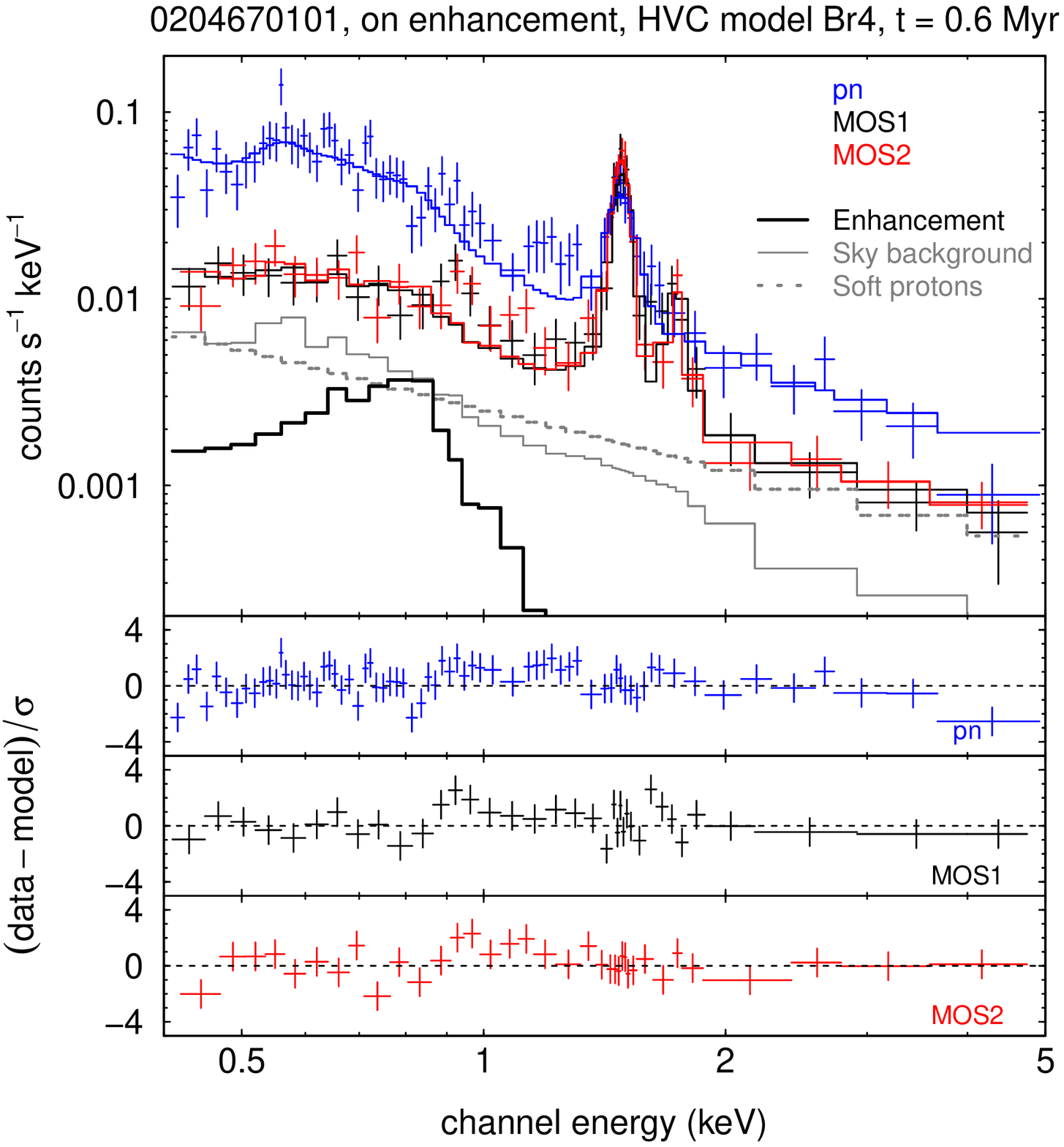}{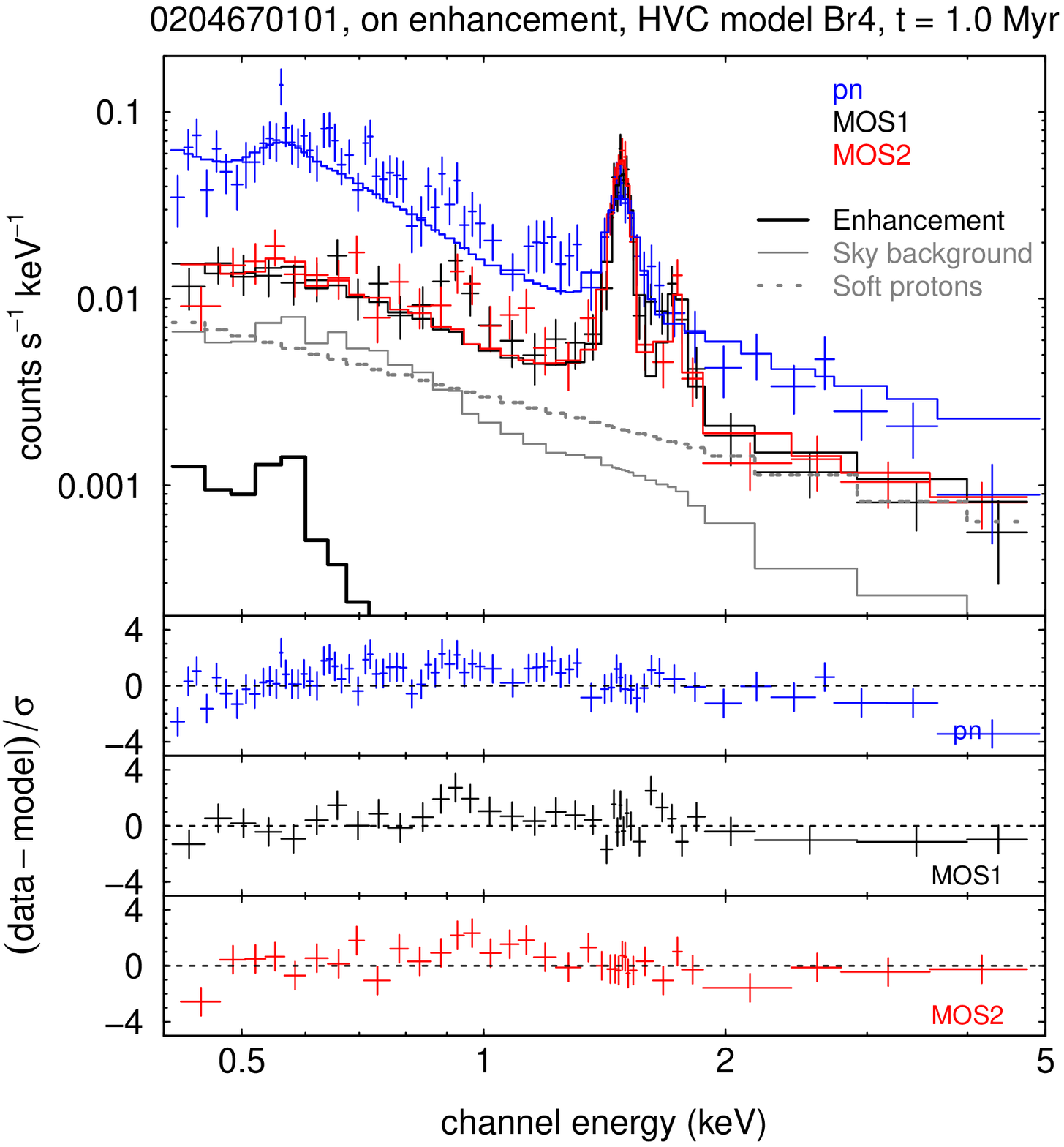}
  \vspace{5mm}
  \plottwo{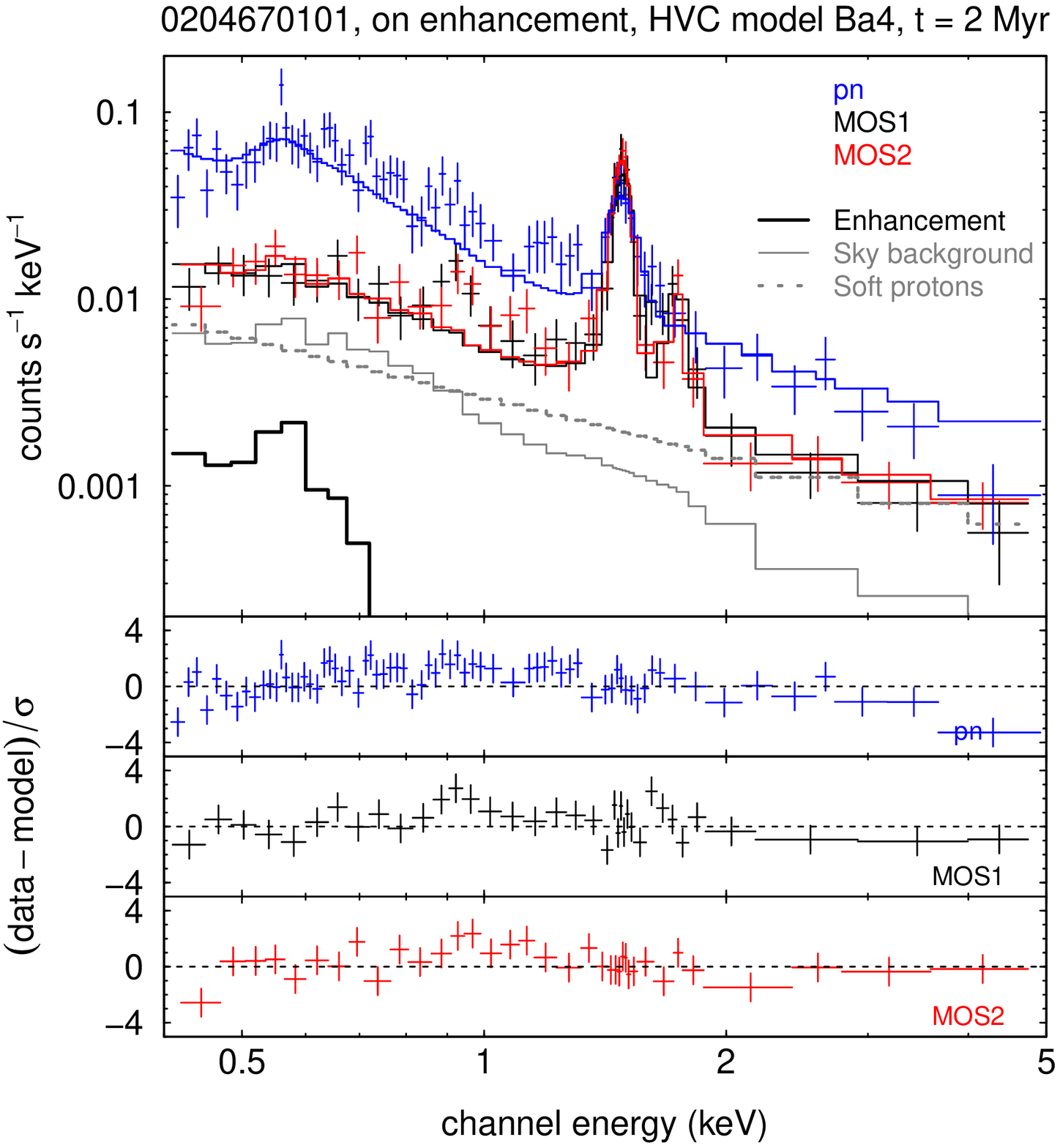}{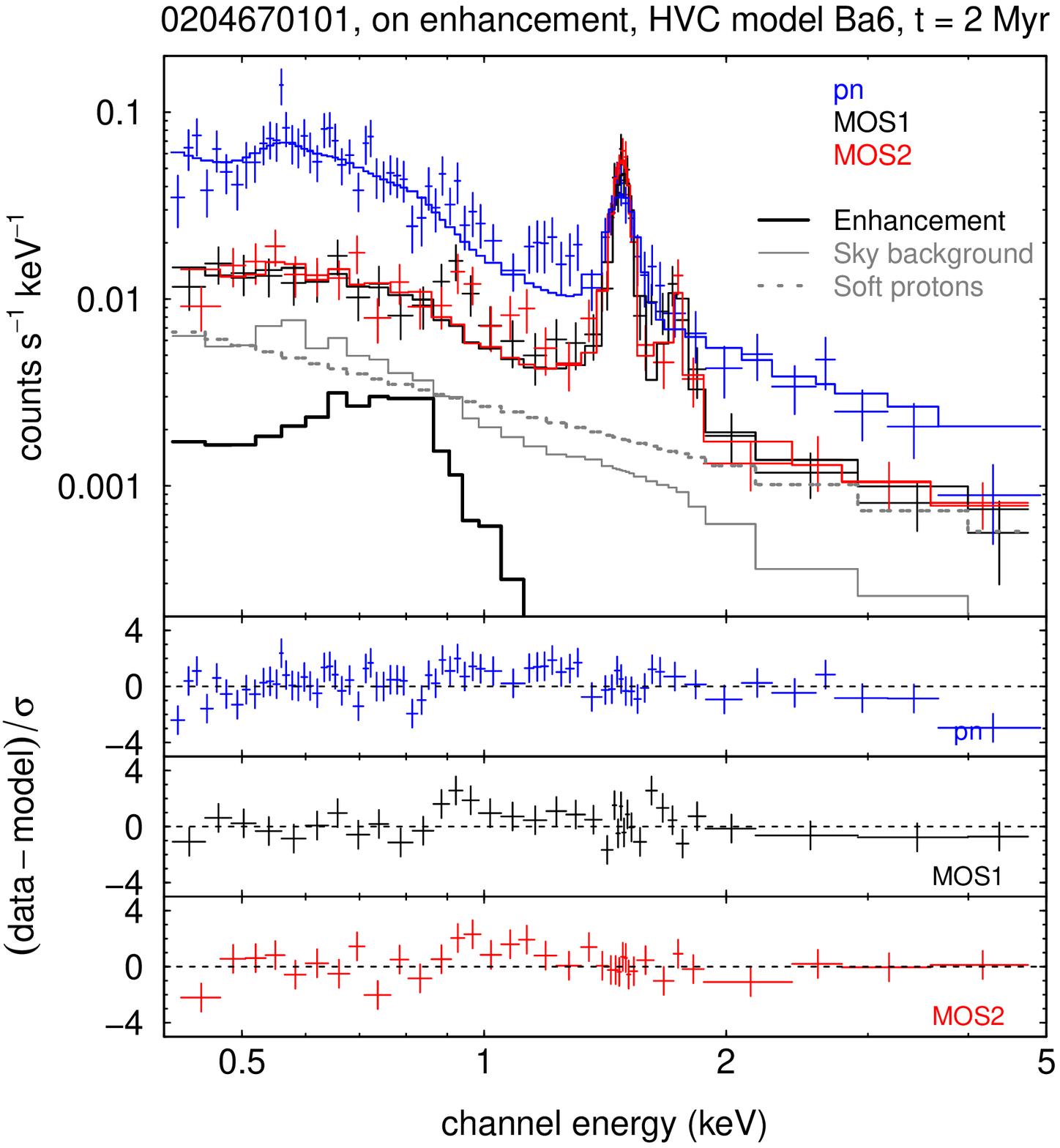}
  \caption{Same as the upper left panel of Figure~\ref{fig:Spectra}, but with the HVC enhancement
    modeled using models from \citet{shelton12} (see plot titles).
    The upper row shows the results for two different epochs of the Br4 model, before and after the
    cloud hits the warm gas. The upper right plot is from the epoch at which the Br4 model emission is
    brightest.
    The lower row shows the results for models Ba4 and Ba6, at $t = 2~\Myr$.
    \label{fig:SpectraHVCModels}}
\end{figure*}

The results are shown in Figure~\ref{fig:HVCresults}. Let us first look at the Br models. The
fastest Br model cloud (Br4, $v = 400~\kmps$) produces enough X-rays to match the measured surface
brightness of the enhancement (upper panel of Figure~\ref{fig:HVCresults}(a)).  This occurs at $t =
1~\Myr$, after the cloud hits the warm gas at $t \sim 0.7~\Myr$ (it takes some time for the
X-ray-emissive gas to build up). However, the resulting X-ray emission is too soft, and the fit to
the observed spectra is poor (lower panel of Figure~\ref{fig:HVCresults}(a), and upper right plot of
Figure~\ref{fig:SpectraHVCModels}). This is because the post-shock temperature expected for a
400~\kmps\ cloud is $\sim$$(\mbox{2.2--2.7}) \times 10^6~\K$ (Equation~(\ref{eq:Postshock}),
assuming that the shock travels up to 10\%\ faster than the cloud), much lower than the measured
temperature of $3.7 \times 10^6~\K$ (for the $1T$ model; Table~\ref{tab:Results}). Higher cloud
speeds would produce higher temperatures, but MS30.7 is unlikely to be exceeding $\sim$400~\kmps. At
earlier epochs, when the cloud is traveling through the hot ambient gas, the post-shock temperatures
are higher ($>$$3.0 \times 10^6~\K$ from Equation~(\ref{eq:Postshock2})), the resulting emission is
harder (upper left plot of Figure~\ref{fig:SpectraHVCModels}), and the fits to the observed spectra
are better (lower panel of Figure~\ref{fig:HVCresults}(a)). However, because of the low density of
the hot gas, the resulting emission is too faint to explain the observations (upper panel of
Figure~\ref{fig:HVCresults}(a); see also discussion in
Section~\ref{subsubsec:ShockHeatingHotMedium}). At later epochs, the X-ray emission fades away, due
to radiative cooling of the hot gas, and the X-ray emission in the model is dominated by that from
the hot background gas. The period of bright X-ray emission lasts less than 1~\Myr.

The slower Br models produce qualitatively the same results as model Br4. However, in these slower
models, the brightest emission is much fainter than the observed emission. Also, this bright
emission fades away much more quickly (in model Br2, the X-ray brightening lasts for $\la$0.1~\Myr).

The Ba models with $v \ge 300~\kmps$ produce about enough, or more than enough X-rays to explain the
MS30.7 emission for a few Myr after hitting the warm gas (upper panel of
Figure~\ref{fig:HVCresults}(b)).  However, only for cloud speeds $\ga$500~\kmps\ is the emission
hard enough to produce reasonable fits to the observed spectra (lower panel of
Figure~\ref{fig:HVCresults}(b), and lower row of Figure~\ref{fig:SpectraHVCModels}).  This is
consistent with the fact that the best-fit $1T$ enhancement temperature implies a shock speed of
$\sim$500~\kmps\ (Section~\ref{subsubsec:StrongShocks}). However, as noted earlier, such a speed is
unreasonably high for MS30.7. Note also that even the hardest HVC model spectrum still underpredicts
the on-enhancement emission around 1~\kev.

The X-ray bright periods in the Ba models last longer than in the Br models, because radiative
cooling is disabled (eventually, adiabatic expansion and cooling causes the X-ray emission to fade
away). In reality, subsolar abundances could suppress the radiative cooling rate, allowing the hot,
X-ray-emissive gas to persist for longer than expected from the Br models. Lowering the abundances
would also tend to lower the emissivity of the hot gas in the \xmm\ band (note that the brightest Ba
models overpredict the X-ray brightness at their earliest epochs; upper panel of
Figure~\ref{fig:HVCresults}(b)).

\subsubsection{Non-equilibrium Ionization}

The model HVC spectra tested above were calculated assuming that the plasma in the hydrodynamical
simulations was in CIE. Similarly, when we used Equations~(\ref{eq:Postshock}) and
(\ref{eq:Postshock2}) to infer shock speeds from the temperatures of the $1T$ and $2T$ enhancement
models, we were assuming that the observed X-ray enhancement is due to emission from a CIE
plasma. In reality, the X-ray emitting plasma may be under- or overionized, relative to the plasma's
electron temperature.

If the plasma is underionized, the resulting spectrum in the \xmm\ band will be softer than that
from a CIE plasma at the same electron temperature. Hence, for reasonable shock speeds, an
underionized plasma would produce emission that is too soft to explain the spectrum of the
enhancement.

In Section~\ref{subsubsec:Recombining}, we used a simple model of an overionized, recombining plasma
to model the X-ray enhancement. This model yielded an electron temperature of $3.2 \times 10^6~\K$,
and an oxygen ionization temperature of $>$$2.8 \times 10^6~\K$, corresponding to shock speeds of 480
and $>$450~\kmps, respectively (Equation~(\ref{eq:Postshock})).\footnote{We are considering only
  strong shocks in a cool or warm medium here, as shocks in a hot medium will likely lead to
  emission that is too faint.} Hence, as with the CIE enhancement models, MS30.7 appears to be
traveling too slowly to explain the observed emission with a recombining plasma model.  Furthermore,
shock-heating typically results in gas that is underionized, rather than overionized. Overionization
typically arises when gas undergoes rapid cooling (via radiation and/or adiabatic expansion),
leaving the high ions frozen in. (These effects can be seen in Figure~4 of \citealt{shelton99}, in
the context of a supernova remnant: at earlier epochs, the shock-heated remnant is underionized,
while at later epochs the cooling remnant is overionized.)

\subsubsection{Summary of Shock Heating}

Shock heating a hot ($\sim$$10^6~\K$), tenuous ($\sim$$10^{-4}~\pcc$) ambient medium results in
emission that is too faint to explain the MS30.7 observations.  In order to adequately model the
MS30.7 X-ray enhancement, we require the cloud to be traveling at $\ga$500~\kmps\ through a cool or
warm ($\la$$10^4~\K$) medium of density $\sim$$10^{-3}~\pcc$. In contrast, MS30.7's speed is
unlikely to exceed $\sim$400~\kmps\ (Section~\ref{subsec:ModelParameters}).  In the context of
MS30.7, the denser warm ambient medium in the \citet{shelton12} models could represent material shed
from a preceding cloud. In this case, the speed of MS30.7 relative to this material is likely to be
even less than the orbital speed of the Magellanic Stream.  We note that higher shock speeds could
in principle be possible if a Galactic wind were impinging upon MS30.7. However, even if such a wind
exists, at the distance of the Magellanic Stream it would likely be too tenuous to produce bright
enough X-ray emission.

The above results therefore rule out simple shock heating as the origin of the MS30.7 X-ray
enhancement. Note that \citet{bregman09a} suggested that a shock driven into the cloud could be
responsible for the X-ray emission. However, \citet{shelton12} found that such a reverse shock would
not heat the cloud to X-ray-emitting temperatures -- in their models, the emission comes from the
shocked ambient medium.

\subsection{Charge Exchange}
\label{subsec:ChargeExchange}

Here we consider the possibility that charge exchange (CX) between neutral atoms in the HVC and ions
in an assumed hot ambient medium is responsible for the observed X-ray emission. This CX emission
will originate in a thin layer of thickness \lCX, where \lCX\ is the mean free path of ions in the
ambient medium undergoing CX with neutrals in the cloud \citep{lallement04b}. This mean free path is
given by
\begin{equation}
  \lCX = \frac{1}{\sigma \ncl} \approx \frac{d}{\sigma \NH},
\end{equation}
where $\sigma$ is the CX cross-section, \ncl\ is the cloud number density, $d$ is the cloud
diameter, and \NH\ is the column density of the cloud. The final part of the above expression
assumes that the extent of the cloud on the sky is similar to that along the line of sight.

CX cross-sections are typically $\sim \mathrm{few} \times 10^{-15}~\cm^2$
\citep[e.g.,][Table~1]{koutroumpa06}, while the column density of the densest part of the cloud is
$\sim$$4 \times 10^{20}~\pcmsq$ (see Figure~\ref{fig:Image}).  Hence, $\lCX /d \sim (\mbox{0.5--1})
\times 10^{-6}$; i.e., the CX emission is expected to arise in a very thin layer around the
cloud. In fact, \lCX\ is an upper limit to the thickness of this layer, since some of the hydrogen
in the cloud may undergo collisional ionization before it is able to undergo CX
\citep{lallement04b}.

If the cloud is traveling at speed $v$ through a hot ambient medium of density \nh, the volumetric
photon emissivity of a specific line, \eCX, due to CX is
\begin{equation}
  \eCX = \sigma y \ncl f A_X \nh v,
\end{equation}
where $y$ is the yield of the line in question, $f$ is the ion fraction for the ion responsible for
the line (e.g., O$^{+8}$ for an \OVIII\ CX line), and $A_X$ is the abundance of the relevant
element.  For a spherical cloud, the emission will arise in a thin hemispherical shell on the upwind
side of the cloud, whose volume is $2 \pi r^2 \lCX$, where $r \approx 50~\pc$ is the cloud radius
(Section~\ref{subsec:ModelParameters}). Hence, if the photon energy for the line in question is $E$,
the total luminosity of the line is
\begin{equation}
  L_\mathrm{line} = 2 \pi r^2 E y f A_X \nh v.
  \label{eq:Lline}
\end{equation}
Note that this luminosity does not depend on the CX cross-section or the cloud density.

Equation~(\ref{eq:Lline}) could be used to calculate a CX spectrum, provided the relevant line
yields and ion fractions for the ambient medium were known. Here we take a different approach, and
estimate the total luminosity due to CX. We introduce an efficiency, $\eta_X$, defined as the
fraction of atoms of element $X$ in the ambient medium that undergo CX reactions with the cloud's
neutrals that lead to the production of X-ray photons within the energy band of interest
(0.4--2.0~\kev, in this case). For example, if oxygen in the ambient medium is mostly O$^{+8}$ and
O$^{+7}$, $\eta_\mathrm{oxygen}$ will be close to 1, as CX will typically result in \OVII\ and
\OVIII\ lines in the band of interest. If oxygen is less highly ionized, $\eta_\mathrm{oxygen}$ will
be smaller, as CX will typically not result in X-ray line emission.

Having defined $\eta_X$, let us define $\mathcal{E}$ as the typical energy of a CX line in the band of
interest.  Then, the total luminosity from a particular element due to charge exchange is
approximately $2 \pi r^2 \mathcal{E} \eta_X A_X \nh v$.  If we sum over all astrophysically abundant elements
that could contribute emission to the band of interest (C, N, O, Ne, Mg, Si, Fe), the total CX
luminosity is
\begin{align}
  L_\mathrm{CX} \approx (8 \times 10^{32}~\ergps)
      \left( \frac{\mathcal{E}}{1~\kev} \right)
      \left( \frac{\sum \eta_X A_X}{1.0 \times 10^{-3}} \right)& \times \nonumber \\
        \left( \frac{Z}{Z_\odot} \right)
      \left( \frac{\phantom{X}\nh\phantom{X}}{10^{-4}~\pcc} \right)
      \left( \frac{\phantom{X}v\phantom{X}}{350~\kmps} \right)&,
  \label{eq:LuminosityCX}
\end{align}
where $Z/Z_\odot$ is the metallicity of the ambient medium relative to solar, and $v = 350~\kmps$ is
the assumed orbital speed of the Magellanic Stream (Section~\ref{subsec:ModelParameters}). Note that
$\sum A_X = 1.0 \times 10^{-3}$ is the sum of the solar abundances of the aforementioned elements
\citep{asplund09}, and so $8 \times 10^{32}~\ergps$ is an upper limit to the charge exchange
luminosity, calculated assuming solar abundances and that $\eta_X = 1$ for each element.

As noted in Section~\ref{subsec:Results}, the intrinsic 0.4--2.0~\kev\ luminosity of the enhancement
is $7.9 \times 10^{33}~\ergps$, 10 times larger than the luminosity given by
Equation~(\ref{eq:LuminosityCX}). (The luminosity of the enhancement is even higher if we use the
$2T$ or recombining models; see Sections~\ref{subsubsec:2T} and \ref{subsubsec:Recombining}.) The
only way CX could be bright enough to explain the observed emission is if the density of the ambient
medium substantially exceeds $10^{-4}~\pcc$, which is unlikely at such a large distance from the
Milky Way (see Section~\ref{subsec:ModelParameters}).

\subsection{Magnetic Reconnection}
\label{subsec:MagneticReconnection}

Finally, we consider magnetic reconnection as a possible source of the X-ray-emitting plasma. In
general, magnetic reconnection occurs when magnetic field lines of different directions move toward
each other. In such an encounter, the topology of the magnetic field can change, and energy stored
in the magnetic field is released via particle acceleration, bulk motion of the plasma, and electric
currents. These currents can then heat the plasma via Ohmic heating.

\citet{zimmer97} studied magnetic reconnection in the context of HVCs interacting with the magnetic
field of the Galaxy. They first estimated the maximum temperature attainable by magnetic
reconnection, by considering the equilibrium between the kinetic energy density of the cloud, the
magnetic energy density, and the thermal energy density in the hot boundary layer in which the
reconnection occurs.  They estimated that temperatures of several million degrees should be
attainable, much higher than the temperatures attainable with shock heating. They confirmed this
estimate with magnetohydrodynamical (MHD) simulations. However, detailed spectral predictions which
we could compare with our \xmm\ observations are unavailable.

Although we wrote above of magnetic reconnection releasing energy stored in the magnetic field and
heating the plasma, ultimately the energy seen in X-rays would come from the kinetic energy of the
cloud. In the magnetic reconnection scenario, the cloud's motion through the Galactic magnetic field
distorts the field, and the distorted field subsequently reconnects. We can place an upper limit on
the rate at which the cloud's kinetic energy can be dissipated in this way by considering the rate
at which the cloud does work against the ambient magnetic pressure, $P_\mathrm{mag} = B^2/8\pi$,
where $B$ is the ambient magnetic field strength. This rate is
\begin{align}
  \Lmag &= \frac{1}{8}B^2 r^2 v \nonumber \\
        &= (1 \times 10^{35}~\ergps)
           \left( \frac{\phantom{X}v\phantom{X}}{350~\kmps} \right)
           \left( \frac{B}{1~\microgauss} \right)^2,
  \label{eq:LuminosityMag}
\end{align}
where we have again used $r = 50~\pc$, $v = 350~\kmps$ is the assumed orbital speed of the
Magellanic Stream, and $B = 1~\microgauss$ is the magnetic field strength estimated assuming
equipartition (Section~\ref{subsec:ModelParameters}). We can also place a lower limit on the time,
\tmag, it would take for the cloud's kinetic energy, \EK, to be completely dissipated via magnetic
reconnection. If we assume that the cloud is spherical, the hydrogen number density is $\nH = \NH /
2r$, and the mass density is $\rho = \NH \mH / 2rX$, where \mH\ is the mass of a hydrogen atom and
$X \approx 0.7$ is the hydrogen mass fraction. Hence, $\EK = \pi r^2 \NH \mH v^2 / 3X$, and
\begin{align}
  \tmag &\equiv \frac{\EK}{\Lmag} \nonumber \\
        &= \frac{8 \pi \NH \mH v}{3 X B^2} \nonumber \\
        &= (9~\Gyr)
           \left( \frac{\phantom{X}v\phantom{X}}{350~\kmps} \right)
           \left( \frac{B}{1~\microgauss} \right)^{-2},
  \label{eq:tmag}
\end{align}
where we have used $\NH = 4 \times 10^{20}~\pcmsq$ (see Figure~\ref{fig:Image}).

The power given by Equation~(\ref{eq:LuminosityMag}) is an order of magnitude larger than the
0.4--2.0~\kev\ luminosity of the enhancement (Section~\ref{subsec:Results}). The time given by
Equation~(\ref{eq:tmag}), meanwhile, is several times the age of the Magellanic Stream
($\sim$1.5--2~\Gyr; e.g., \citealt{gardiner96,nidever08}). It therefore seems that, from an
energetics point of view, magnetic reconnection could plausibly power the observed X-ray emission
from MS30.7.

\section{DISCUSSION}
\label{sec:Discussion}

\subsection{Is the Emission Really from MS30.7?}
\label{subsec:IsEmissionFromMS30.7}

Before we discuss our results, let us first consider again the possibility that the observed X-ray
emission is not physically associated with MS30.7, but is the result of a chance alignment. We first
considered this issue in Section~\ref{subsec:ImageCreation} -- although we could not confidently
rule out such a chance alignment, the morphology of the enhancement argues in favor of its being
associated with MS30.7. Here, we specifically consider the possibility that the observed X-ray
emission is due to a chance alignment with a background group of galaxies. Such objects exhibit
X-ray temperatures of $\sim$$10^7~\K$ \citep{osmond04}, similar to that of the hotter component of
our $2T$ enhancement model (Section~\ref{subsubsec:2T}).

From the NASA Extragalactic Database (NED\footnote{http://ned.ipac.caltech.edu/}), we find five
galaxy groups within our mosaicked \xmm\ field of view. One of these, MZ~01537, is centered at
$(\alpha,\delta) = (00^\mathrm{h}13^\mathrm{m}38\fs1, -27\degr10\arcmin51\arcsec)$, on the eastern
edge of the X-ray enhancement. This group is at a redshift of $z = 0.1264$.  The size of this group
is not stated, but assuming a typical group radius of $\sim$0.5~\Mpc\ \citep{osmond04}, the radius
on the sky is $\sim$3.2\arcmin\ (from NED, calculated using the five-year \textit{WMAP} cosmology
parameters; \citealt{komatsu09}). The position and estimated size of MZ~01537 is shown by the dashed
yellow circle in the lower panel of Figure~\ref{fig:Image}.

The observed X-ray enhancement is located in the western half of the galaxy group, and beyond the
group's estimated western edge. If the enhanced X-ray emission were from this galaxy group, we would
expect the emission to be centered on the group's center. We therefore conclude that the X-ray
enhancement is not associated with a background galaxy group.

\subsection{Comparison with \citet{bregman09a}}

As stated in the Introduction, \citet{bregman09a} reported an enhanced 0.4--1.0~\kev\ pn count rate
of $0.64 \pm 0.10$ counts \pks\ \parcminsq\ toward the densest part of MS30.7. They obtained this
value by extracting count rates from twenty equal regions around an annulus centered on the peak of
the exposure map, with inner and outer radii of $\approx$4.7\arcmin\ and $\approx$10.5\arcmin. Since
this annulus was centered on the peak of the exposure map, the camera sensitivity was the same in
all twenty regions, and so differences in the count rate correspond directly to differences in the
observed X-ray surface brightness. \citet{bregman09a} obtained an on-cloud count rate of $2.54 \pm
0.09$ counts \pks\ \parcminsq\ from the three regions toward the densest part of the cloud, and a
background count rate of $1.90 \pm 0.05$ counts \pks\ \parcminsq\ from six regions to the south-west
of the cloud. The difference between these two rates yields the count rate for the on-cloud
enhancement quoted above.

From our best-fit $1T$ model of the enhancement, the 0.4--1.0~\kev\ intrinsic surface brightness is
$6.7 \times 10^{-16}~\flux\ \parcminsq$. We used
PIMMS\footnote{http://cxc.harvard.edu/toolkit/pimms.jsp} to convert this to a count rate for the
\xmm\ pn camera with the thin filter, obtaining $0.64$ counts \pks\ \parcminsq.\footnote{For this
  conversion we used $\NH = 1.6 \times 10^{20}~\pcmsq$ (Section~\ref{subsec:Method}) and $\log T =
  6.55$ (the nearest value to our measured temperature; Table~\ref{tab:Results}).} Although this
number is in exact agreement with \citeauthor{bregman09a}'s value, it should be noted that the PIMMS
flux-to-count rate conversion is for an on-axis point source. As described above, \citet{bregman09a}
extracted their count rates from within an annular region centered on the peak of the exposure
map. From the 0.4--1.0~\kev\ pn exposure map, we find that the average sensitivity within this
annular region is $\sim$60\%\ of the peak sensitivity. Hence, the count rate inferred from our
best-fit $1T$ model of the enhancement is $\sim$60\%\ of the value quoted by \citet{bregman09a}.
This discrepancy may indicate that our best-fit $1T$ model does not capture all of the soft X-ray
emission from the enhancement, or that the soft-proton contamination was not uniform over the pn
detector during \citeauthor{bregman09a}'s observation.

\citet{bregman09a} state that their count rate measurement corresponds to a luminosity of $4 \times
10^{33}~\ergps$, which is about half of the 0.4--1.0~\kev\ luminosity derived from our best-fit $1T$
enhancement model, $7.2 \times 10^{34}~\ergps$ (note that this is only $\sim$10\% less than the
0.4--2.0~\kev\ luminosity reported in Section~\ref{subsec:Results}, as our best-fit $1T$ enhancement
model produces little emission above 1~\kev\ (see Figure~\ref{fig:Spectra})). It is unclear how this
discrepancy in the luminosities arises, given that our best-fit model yields a count rate that is
smaller than the value reported by \citet{bregman09a}.

\subsection{The Origin of the X-ray Emission}
\label{subsec:Origin}

We examined models for the origin of the X-ray emission in Section~\ref{sec:Models}.  We found that
neither turbulent mixing with or compression of a hot ambient medium
(Section~\ref{subsec:TurbulentMixing}), shock heating (Section~\ref{subsec:ShockHeating}), nor CX
(Section~\ref{subsec:ChargeExchange}) can adequately explain the observed emission. Strong shocks in
a cool or warm ambient medium result in emission that is too soft (for reasonable cloud speeds),
while turbulent mixing, compression or shock heating of a hot ambient medium, and CX all result in
emission that is too faint (for reasonable ambient densities).

Although we do not have spectral predictions that we can directly compare with our observations,
magnetic reconnection appears to be the best explanation for the observed emission. \citet{zimmer97}
found that this process could heat plasma to temperatures of several million degrees (much hotter
than is possible with shock heating), and we argued that, from an energetics point of view, the
resulting emission could be bright enough to match the observations
(Section~\ref{subsec:MagneticReconnection}).  However, if the magnetic field strength in the
vicinity of the Magellanic Stream is substantially less than 1~\microgauss, or if the efficiency
with which the cloud's kinetic energy can be converted to thermal energy in the X-ray-emitting
plasma is $\la$0.1, then magnetic reconnection also has difficulty explaining the observed X-ray
emission. Resistive MHD simulations are needed to determine the X-ray spectrum and brightness
that would result from a MS30.7-like cloud interacting with the Galaxy's magnetic field. Such
simulations would have to take into account the subsolar metallicity of the Magellanic Stream
\citep{fox13}.

In Section~\ref{sec:Models}, we concentrated on the spectrum and brightness of the observed
emission.  Let us conclude this discussion by considering the morphology of the emission.  The
Magellanic Stream in general (and MS30.7 in particular) is likely moving in the general direction of
the Magellanic Clouds, which lie south to south-east of MS30.7. The X-ray emission is mainly to the
north and west of the densest part of the cloud, i.e., on the downstream side of the cloud. In
contrast, shock-heated gas is expected to be on the upstream side of the cloud
\citep{shelton12}. Similarly, we pointed out in Section~\ref{subsec:ChargeExchange} that CX emission
is expected to originate in a thin shell on the upstream side of the cloud.

\citet{zimmer97} suggested that magnetic reconnection would take place throughout the mixing layer
between an HVC and the ambient medium, as the fluid flow in this mixing layer would tend to tangle
up the field, bringing oppositely directed magnetic fields together at many different locations. In
this scenario, we would expect to see X-ray emission all around the cloud, or possibly concentrated
on the upstream side of the cloud, rather than concentrated on the downstream side.  However, we
suggest that it may be possible for magnetic reconnection to preferentially heat the plasma on the
downstream side of the cloud. As an HVC moves through a magnetic field roughly perpendicular to its
velocity vector, the field is drawn down into a ``V'' shape behind the cloud. This can be seen in
two-dimensional MHD simulations of HVCs (\citealt{santillan99}, Figures~4--6; \citealt{jelinek11},
Figure~4), and is also reported to be seen in three-dimensional simulations \citep{kwak09}. If this
V shape is sufficiently deep and steep-sided, the downward-pointing magnetic field on one side of
the V will be adjacent to the upward-pointing field on the other side. Such a field configuration
could allow magnetic reconnection to occur behind the cloud. Resistive MHD simulations would be
needed to test whether or not this would occur in practice, i.e., do the oppositely directed
magnetic fields in the V get close enough to each other for reconnection to take place?

\section{SUMMARY AND CONCLUSIONS}
\label{sec:Summary}

We have presented our analysis of two \xmm\ observations of the HVC MS30.7$-$81.4$-$118 (MS30.7), a
constituent of the Magellanic Stream. We concentrated on the enhanced X-ray emission observed near
the densest part of the cloud, initially reported by \citet{bregman09a}. This enhanced emission is
concentrated to the north and west of the densest part of the cloud -- this is likely the downstream
side of the cloud. The X-ray enhancement is $\sim$6\arcmin\ or $\sim$100~\pc\ across
(Section~\ref{subsec:ImageCreation}).

We first modeled the enhancement with a $1T$ thermal plasma model, obtaining a temperature of $3.7
\times 10^6~\K$, and an intrinsic 0.4--2.0~\kev\ luminosity of $7.9 \times 10^{33}~\ergps$
(Section~\ref{subsec:Results}).  While the fit was reasonably good overall, the on-enhancement
emission around 1~\kev\ tended to be underpredicted. The fit could be improved by the addition of a
second plasma component -- in this $2T$ fit, the best-fit temperatures were $2.6 \times 10^6$ and
$1.2 \times 10^7~\K$, and the luminosity was $\sim$70\% larger than that obtained with the $1T$
model (Section~\ref{subsubsec:2T}). The fit could also be improved by modeling the enhancement as a
recombining plasma (Section~\ref{subsubsec:Recombining}). However, the fact that this model yields a
nitrogen ionization temperature lower than the electron temperature ($<$$1.3 \times 10^6$ versus
$3.2 \times 10^6~\K$) suggests that, if recombination emission really is important, our recombining
spectral model may be overly simplistic.

We examined several different physical models for the observed X-ray emission
(Section~\ref{sec:Models}; see also Section~\ref{subsec:Origin}).  Shock heating of hot, tenuous gas
and of warm, denser gas results in emission that is too faint and too soft, respectively, while
turbulent mixing, compression of a hot ambient medium, and CX all result in emission that is too
faint. Magnetic reconnection appears to be the best explanation for the heating of the
X-ray-emitting gas. Resistive MHD simulations are needed to test this conclusion. In particular,
does magnetic reconnection dissipate the cloud's kinetic energy with sufficient efficiency to power
the observed X-ray emission, and is the resulting emission concentrated on the downstream side of
the cloud? If such simulations can explain the observed X-ray emission, then, as noted in the
Introduction, the observed X-ray spectrum and brightness could potentially constrain the magnetic
field in the vicinity of the Magellanic Stream.

We conclude by noting that there is no reason to think that MS30.7 is special. Other similar
constituent clouds of the Magellanic Stream may exhibit X-ray emission. If magnetic reconnection is
indeed responsible for the emission from MS30.7, and if the ambient conditions are reasonably
uniform all along the Magellanic Stream, we might expect other clouds to be similar to MS30.7 in
terms of the inferred plasma temperature, the X-ray luminosity, and the location of the
X-ray-emitting plasma. This expectation could be tested with future observations.

\acknowledgements

We thank C. Br{\"u}ns for providing us with the \HI\ data shown in Figure~\ref{fig:Image}.
This research is based on observations obtained with \xmm, an ESA science mission with instruments
and contributions directly funded by ESA Member States and NASA.
We acknowledge use of the R software package \citep{R}.
This research has made use of the NASA/IPAC Extragalactic Database (NED) which is operated by the
Jet Propulsion Laboratory, California Institute of Technology, under contract with the National
Aeronautics and Space Administration.
This research was funded by NASA grant NNX12AI56G, awarded through the Astrophysics Data Analysis
Program.
K.K. was supported by the year 2013 Research Fund of the Ulsan National Institute of Science and
Technology (UNIST).

\bibliography{references}

\end{document}

%% file: newcom.tex
\newcommand{\CIV}{\ion{C}{4}}

\newcommand{\HI}{\ion{H}{1}}

\newcommand{\NV}{\ion{N}{5}}

\newcommand{\NeIX}{\ion{Ne}{9}}
\newcommand{\OVI}{\ion{O}{6}}
\newcommand{\OVII}{\ion{O}{7}}
\newcommand{\OVIII}{\ion{O}{8}}

\newcommand{\Kalpha}{K$\alpha$}

\newcommand{\mH}{\ensuremath{m_{\mathrm{H}}}}

\newcommand{\Ne}{\ensuremath{n_{\mathrm{e}}}}

\newcommand{\nH}{\ensuremath{n_{\mathrm{H}}}}

\newcommand{\NH}{\ensuremath{N_{\mathrm{H}}}}

\newcommand{\nm}{\ensuremath{\mbox{\nm}}}
\newcommand{\cm}{\ensuremath{\mbox{cm}}}

\newcommand{\km}{\ensuremath{\mbox{km}}}

\newcommand{\pc}{\ensuremath{\mbox{pc}}}
\newcommand{\kpc}{\ensuremath{\mbox{kpc}}}
\newcommand{\Mpc}{\ensuremath{\mbox{Mpc}}}
\newcommand{\s}{\ensuremath{\mbox{s}}}
\newcommand{\ks}{\ensuremath{\mbox{ks}}}

\newcommand{\Myr}{\ensuremath{\mbox{Myr}}}
\newcommand{\Gyr}{\ensuremath{\mbox{Gyr}}}
\newcommand{\gram}{\ensuremath{\mbox{g}}}

\newcommand{\kev}{\ensuremath{\mbox{keV}}}

\newcommand{\erg}{\ensuremath{\mbox{erg}}}

\newcommand{\microgauss}{\ensuremath{\mu\mbox{G}}}

\newcommand{\K}{\ensuremath{\mbox{K}}}

\newcommand{\counts}{\ensuremath{\mbox{counts}}}

\newcommand{\parcminsq}{\ensuremath{\mbox{arcmin}^{-2}}}
\newcommand{\pdegsq}{\ensuremath{\mbox{deg}^{-2}}}
\newcommand{\pcc}{\ensuremath{\cm^{-3}}}
\newcommand{\pcmsq}{\ensuremath{\cm^{-2}}}

\newcommand{\pks}{\ensuremath{\ks^{-1}}}
\newcommand{\ps}{\ensuremath{\s^{-1}}}

\newcommand{\dm}{\ensuremath{\pcc\ \pc}}
\newcommand{\emismeas}{\ensuremath{\cm^{-6}}\ \pc}
\newcommand{\ergps}{\erg\ \ps}
\newcommand{\flux}{\erg\ \pcmsq\ \ps}

\newcommand{\kmps}{\km\ \ps}

\newcommand{\chandra}{\textit{Chandra}}

\newcommand{\rosat}{\textit{ROSAT}}

\newcommand{\xmm}{\textit{XMM-Newton}}

\newcommand{\citepossessive}[1]{\citeauthor{#1}'s \citeyearpar{#1}}

\providecommand{\eqref}[1]{equation~(\ref{#1})}

\newcommand{\esas}{\textit{XMM}-ESAS}